\documentclass[sigconf, screen]{acmart}
\AtBeginDocument{%
  }

\setcopyright{none}
\copyrightyear{2025}
\acmYear{2025}
\acmDOI{XXXXXXX.XXXXXXX}
\acmConference[XXXX '25]{XXXX}{XXX XX--XX,
  2025}{XXX, XX, USA.}

\acmISBN{XXX-X-XXXX-XXXX-X/XXXX/XX}

\usepackage{amsmath, amsthm}
    
\usepackage{graphicx, xcolor} % xcolor instead of color
\usepackage{hyperref, url}
\usepackage{array}
\usepackage{booktabs, multirow}
\usepackage{caption}
\usepackage{subcaption}
\usepackage{enumitem}
\usepackage{setspace}
\usepackage{multicol}
\usepackage{multirow}
\usepackage[ruled, linesnumbered, norelsize]{algorithm2e}

\usepackage{tikz}
\newcommand{\redcircled}[1]{%
  \tikz[baseline=(char.base)]{
    \node[shape=circle, fill=red, text=white, inner sep=1pt] (char) {#1};
  }%
}
\newcommand{\blackcircled}[1]{%
  \tikz[baseline=(char.base)]{
    \node[shape=circle, fill=black, text=white, inner sep=1pt] (char) {#1};
  }%
}
\begin{document}

\title{KVComp: A High-Performance, LLM-Aware, Lossy Compression Framework for KV Cache}
% \subtitle{\normalsize{SC 2025 Submission
    % \textbf{\#xxx}}}
\author{Bo Jiang}
\affiliation{
  \institution{Temple University}
  \city{Philadelphia}
  \state{PA}
  \country{USA}
}
\email{jiang.bo@temple.edu}

\author{Taolue Yang}
\affiliation{
  \institution{Temple University}
  \city{Philadelphia}
  \state{PA}
  \country{USA}
}
\email{taolue.yang@temple.edu}

\author{Youyuan Liu}
\affiliation{
  \institution{Temple University}
  \city{Philadelphia}
  \state{PA}
  \country{USA}
}
\email{youyuan.li@temple.edu}

\author{Chengming Zhang}
\affiliation{
  \institution{University of Houston}
  \city{Houston}
  \state{TX}
  \country{USA}
}
\email{czhang59@Central.UH.EDU}

\author{Xubin He}
\affiliation{
  \institution{Temple University}
  \city{Philadelphia}
  \state{PA}
  \country{USA}
}
\email{xubin.he@temple.edu}

\author{Sian Jin}
\affiliation{
  \institution{Temple University}
  \city{Philadelphia}
  \state{PA}
  \country{USA}
}
\email{sian.jin@temple.edu}

\begin{abstract}
  % Lossy compression is one of the most effective methods for reducing the size of scientific data containing multiple data fields.
  % It reduces information density through prediction or transformation techniques to compress the data. 
  % Previous approaches use local information from a single target field when predicting target data points, limiting their potential to achieve higher compression ratios.
  % In this paper, we identified significant cross-field correlations within scientific datasets. We propose a novel hybrid prediction model that utilizes CNN to extract cross-field information and combine it with existing local field information. Our solution enhances the prediction accuracy of lossy compressors, leading to improved compression ratios without compromising data quality. We evaluate our solution on three scientific datasets, demonstrating its ability to improve compression ratios by up to 25\% under specific error bounds. Additionally, our solution preserves more data details and reduces artifacts compared to baseline approaches. 
Transformer-based large language models (LLMs) demonstrate impressive potential in various practical applications. However, long context inference poses a significant challenge due to the enormous memory requirements of the key-value (KV) cache, which can scale to multiple gigabytes as sequence length and batch size increase. In this paper, we present KVComp, a generic and efficient KV cache management framework optimized for long-text generation that synergistically works with both latency-critical and throughput-critical inference systems. KVComp employs novel lossy compression techniques specifically designed for KV cache data characteristics, featuring careful co-design of compression algorithms and system architecture. Our approach maintains compatibility with the growing nature of KV cache while preserving high computational efficiency. Experimental results show that KVComp achieves on average 47\% and up to 83\% higher memory reduction rate compared to existing methods with little/no model accuracy degradation.
Furthermore, KVComp achieves extremely high execution throughput, effectively reducing decompression overhead and, in some cases, even accelerating the matrix-vector multiplication operation and outperform cuBLAS-based attention kernels with less data movement.

% In addition, matrix-vector multiplication benchmarks demonstrate that KVComp achieves comparable or even better efficiency than PyTorch’s cuBLAS in some cases.

% Furthermore, when compared to state-of-the-art KV cache quantization approaches, KVComp delivers better memory savings with zero accuracy degradation, requiring only minimal additional computational overhead.
  % Lossy compression is one of the most popular and effective methods for data reduction in scientific computing. It reduces information entropy by using prediction or transformation techniques, then to achieve a higher compression ratio. Previous compressors use local information from the compression target field to make prediction, ignoring the potential of cross-field correlations and limiting their ability to achieve high compression ratios. In this paper, we propose a novel Neural network-based hybrid prediction model which leverages both the local information and the cross-field information to make a more accurate prediction. Our solution enhances the prediction accuracy of existing prediction-based lossy compressors, and achieves a higher compression ratio without compromising data quality. When the data quality requirement is relative low, we can even only use the cross-field information for prediction to achieve an extremely high compression ratio. We evaluation our solution on three scientific datasets. It improves the compression ratio by up to more than 90\% on some fields and also improves the overall compression ratio of the whole dataset.
\end{abstract}





\maketitle

\section{Introduction}
\label{sec:introduction}

\textcolor{black}{Transformer-based large language models (LLMs) have revolutionized natural language processing, enabling breakthroughs in diverse tasks\cite{brown2020language, taylor2022galactica}.}
The self-attention mechanism allows models to capture long-range dependencies and contextual information. 
However, these capabilities come at a significant computational and memory cost during inference with long input contexts, \textcolor{black}{where the memory footprint of the key-value (KV) cache becomes a major bottleneck\cite{kwon2023efficient, pope2023efficiently}.}

The KV cache stores intermediate key and value tensors for each token processed by the model and is reused during subsequent decoding steps to avoid redundant computation. As sequence length and batch size increase, the cache size grows linearly and can consume a substantial portion of GPU memory, \textcolor{black}{sometimes exceeding the memory footprint of the model weights themselves\cite{liu2023scissorhands}.}
For example, LLaMA2-30B inference with a context length of 32,000 and a batch size of 8 can produce over 100 GB of KV cache, surpassing the model size itself (i.e., 60 GB in float16).
This growing footprint severely constrains the inference performance, \textcolor{black}{limiting the achievable context length, reducing batch size, or impeding the deployment of LLMs on memory-constrained hardware\cite{kwon2023efficient, liu2023scissorhands}.}

To address these challenges, recent studies mainly leverage three approaches to reduce KV cache size: quantization, pruning, and GPU-CPU migration. 
\textcolor{black}{Quantization-based methods\cite{liu2024kivi, hooper2024kvquant}}, such as KIVI, aim to carefully design quantization strategies that minimize the impact on model accuracy. 
However, these techniques often achieve modest compression ratios unless combined with additional encoding, which introduces overhead and limits their applicability in latency-sensitive LLM inference. 
\textcolor{black}{Pruning-based methods\cite{wang2021spatten, zhang2024q}}, such as Q-Hitter, selectively discard KV pairs that are predicted to be unimportant for future decoding. 
While effective in some cases, these methods can suffer from unpredictable attention callbacks, leading to either costly KV recomputation or substantial accuracy degradation. 
\textcolor{black}{GPU-CPU migration is a traditional approach for handling memory overflows, offloads KV cache data to CPU memory\cite{sheng2023flexgen}.} Although this mitigates GPU memory pressure, it significantly degrades inference performance due to data transfer latency and complex scheduling overhead.
% quantization techniques to reduce KV cache size while preserving model accuracy.
% For example, KIVI propose to quantize the KV cache in various direction and granularity.
% However, these approaches often achieve limited compression ratios unless paired with additional encoding, which introduces non-negligible overhead and compromises their practicality in latency-sensitive LLM inference. Another solution is GPU-CPU migration, or offloading, which shifts KV cache data to CPU memory to reduce GPU usage. While this approach alleviates memory constraints, it incurs significant communication overhead and requires complex scheduling, making it unsuitable for high-throughput or real-time applications.

In this paper, we present KVComp, a high-performance, LLM-aware lossy compression framework tailored for KV cache during inference. 
KVComp combines error-controlled quantization with GPU-based high-throughput entropy encoding and cache-resident decompression to deliver significant memory savings while preserving computational efficiency. 
By co-designing the compression pipeline and system-level execution, KVComp enables decompressed data to be consumed in situ within GPU shared memory or registers, thereby avoiding global memory writeback and maximizing throughput.
Our high-throughput design makes KVComp capable for deployment in LLM inference with minimum overhead.
Note that our approach is orthogonal to existing pruning techniques and GPU-CPU migration strategies.

Our key contributions are as follows:
\begin{enumerate}
    \item    We introduce a system-aware lossy compression pipeline based on a 2D blockwise design, combining fine-grained quantization with GPU-efficient Huffman encoding to strike a balance between compression ratio and model accuracy.
    \item  We implement a cache-resident decompression solution, which is high-throughput, branch-divergence-free decompression method that fuses decoding with matrix-vector multiplication, eliminating unnecessary memory transfers.
    \item  We integrate KVComp with three LLMs and evaluate its accuracy across two benchmarks for LLM inference. Our results demonstrate up to 83\% improvement in compression ratio over state-of-the-art quantization methods, with negligible or no degradation in accuracy. Moreover, KVComp achieves exceptionally high execution speed, effectively hiding decompression overhead and, in some cases, even accelerating the matrix-vector multiplication operation to outperform cuBLAS-based attention kernels.
\end{enumerate}

The rest of this paper is organized as follows: Section 2 provides a background on LLM inference, lossy compression, and GPU architecture. Section 3 presents the design of our KVComp framework. Section 4 details our experimental evaluation and analysis. Lastly, Section 5 concludes the paper and future research directions.
\section{Background and Motivation}
\label{sec:background}

\subsection{Large Language Model Inference}

LLMs have become foundational in natural language processing, \textcolor{black}{demonstrating remarkable capabilities in tasks such as text generation\cite{achiam2023gpt}.} The inference process for these models, particularly for generating long sequences of text (i.e., long context inference), typically involves auto-regressive decoding. \textcolor{black}{In this process, the model generates output tokens one by one, with each new token depending on the previously generated ones and the initial input prompt\cite{radford2018improving, vaswani2017attention}.}

\textcolor{black}{The core component of the Transformer architecture enabling this process is the attention mechanism.\cite{vaswani2017attention}}
% \cred{xh: citations needed here}. 
During inference, intermediate states known as the Key (K) and Value (V) tensors are computed for each token within the self-attention layers. 
\textcolor{black}{These K and V tensors can be reused for processing future tokens and are collectively referred to as the KV cache\cite{shazeer2019fast}.}
% These K and V tensors, collectively referred to as the KV cache, store context computed from previous tokens. 
This cache is crucial because it allows the model to efficiently attend to relevant parts of the preceding sequence without recomputing these states for every new token generation step. 
The KV cache typically has a structure represented as $[context\_len, head\_num, head\_dim]$, where $context\_len$ is the sequence length so far, $head\_num$ is the number of attention heads, and $head\_dim$ is the dimension of each head.
% \cred{xh: should we use a figure to show the KV cache workflow in the context of LLM?}\cb{added 1a and 1b}

\begin{figure}[t]
  \centering
  \begin{subfigure}[b]{0.45\textwidth}
        \centering
        \includegraphics[width=\textwidth]{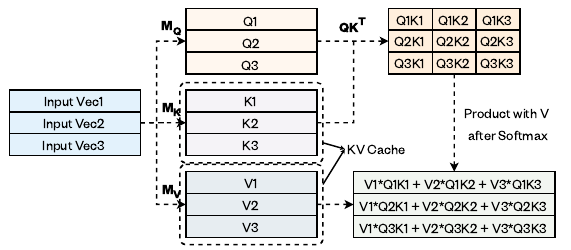}
        \caption{KV cache in prefill stage(assuming 3 token input)}
        \label{fig:kv_cache_prefill}
    \end{subfigure}
    \hfill
    \begin{subfigure}[b]{0.45\textwidth}
        \centering
        \includegraphics[width=\textwidth]{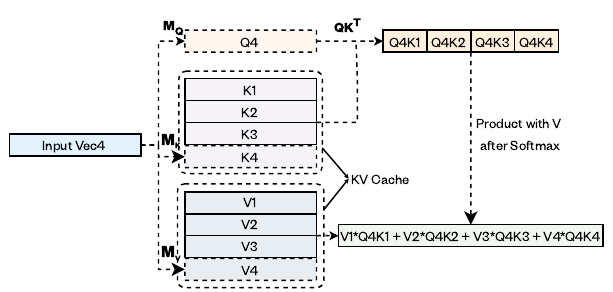}
        \caption{KV cache in decode stage}
        \label{fig:kv_cache_decode}
    \end{subfigure}
    \vspace{-0.2cm}
    \caption{KV Cache behavior during LLM Inference, \(M_{Q,K,V}\) is the mapping matrix to \(K,Q,V\) vector. Each \(Q\) vector perform dot product with every \(K\) vector to generate a \textit{weight} after softmax operation for each \(V\) vector. All the \(V\) vectors multiplied with their \textit{weights} and aggregate to a output vector \textit{Attn Output}.}
    \Description{Two diagrams showing KV cache operations: left shows prefill stage with parallel processing of 3 input tokens, right shows decode stage with sequential token generation. Both include attention mechanism components with Q, K, V vectors and matrix operations.}
    \vspace{-0.5cm}
\end{figure}

\textcolor{black}{KV cache mainly participate in two stages\cite{agrawal2023sarathi} during inference:}
% LLM inference generally proceeds in two main stages:
\textbf{Prefill Stage:}  The initial user prompt is processed in parallel by the model. This stage generates the initial KV cache based on the prompt. Shown in figure~\ref{fig:kv_cache_prefill}.
\textbf{Decode Stage:}  The model generates subsequent tokens auto-regressively. In each step, a new token is produced, its corresponding KV vectors are calculated and appended to the existing KV cache, and the entire updated cache will be used to generate the next token. This stage typically involves matrix-vector multiplications using the KV cache within the attention mechanism. Shown in figure~\ref{fig:kv_cache_decode}.

The KV cache is essential for avoiding redundant computation and allows efficient inference within a reasonable response time. 
However, it also introduces high memory consumption.
% While essential for performance, the KV cache presents a significant challenge: memory consumption. 
\textcolor{black}{As the generated sequence length $context\_len$ and batch size increase, the KV cache grows linearly in size, potentially consuming as much or even more GPU memory than the model weights themselves\cite{kwon2023efficient, liu2023scissorhands}.}
% As the generated sequence length $context\_len$ and the batch size increase, the size of the KV cache grows linearly, potentially consuming dozens of gigabytes of GPU global memory.
% \cred{xh: why is this a big challenge? From the system aspect, consuming GBs of memory is not a concern considering today's computers can easily have 100+ GB memory}\cb{bo: The inference happens in GPU so KV cache usually store in GPU global memory. Most of the GPU memory are occupied by model weights, only several gigabytes left for KV cache} 
This large memory footprint can become a major bottleneck, limiting the \textcolor{black}{maximum sequence length and batch size, degrading inference performance, or even preventing inference on hardware with constrained memory resources\cite{kwon2023efficient, liu2023scissorhands}. }
This memory challenge motivates the development of techniques to manage the KV cache efficiently.
\textcolor{black}{KV cache quantization has been widely explored in recent studies\cite{liu2024kivi, hooper2024kvquant}.}
However, without high-speed entropy-based encoding, quantization alone offers only limited data reduction to preserve accuracy.
\textcolor{black}{GPU-CPU data migration, also known as offloading, is another approach to reduce GPU memory consumption by utilizing the CPU memory pool\cite{sheng2023flexgen}.} 
While it mitigates the risk of GPU memory overflow, it introduces significant performance overhead due to data transfer latency and complex scheduling.
% This memory challenge motivates the development of techniques to manage the KV cache efficiently, such as the quantization, data migration, and compression methods explored in this paper. Optimizing the storage and access of the KV cache, particularly during the decode stage where it is repeatedly accessed, is critical for enabling efficient and scalable long-context LLM inference.

\begin{figure*}[tbp]
    \centering
   \includegraphics[width=0.85\linewidth]{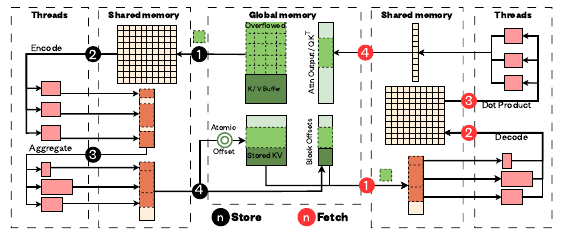}
   % \vspace{-0.3cm}
    \caption{Overview of KVComp. \textit{Store (black nodes)}: 1. Load truncated 2d block to shared memory. 2. Launch threads to encode the vector for each KV vector, write the result to the thread's buffer. 3. Aggregate all encoded bits to a whole chunk of bits. 4. Update the atomic buffer offset, write the chunk to the compressed data buffer and the offset of the chunk to the block offsets array. \textit{Fetch (red nodes)}: 1\&2. Reverse operation of 3 and 4 in store. 3. With the decoded 2d block in shared memory, each thread takes responsibility for one dot product and write the result back to global memory for LLM Inference. 4. Efficiently write matrix vector multiplication result back to GPU global memory.} 
    \Description{Flowchart diagram showing KVComp system architecture with two main paths: Store path (black nodes) for compression and Fetch path (red nodes) for decompression. Includes components like KV cache input, 2D block processing, shared memory operations, and final output to LLM inference.}
    \label{fig:overview}
    \vspace{-0.3cm}
\end{figure*}

\subsection{Lossy Compression}

% Lossy compression aim to reduce the data size by discarding information deemed less critical, accepting a certain degree of data fidelity loss in exchange for a higher compression ratio compared to lossless methods. 
% This approach is particularly relevant in scenarios where exact data reconstruction is not strictly necessary while significant memory or bandwidth savings are paramount.

In large neural networks, particularly LLMs, \textcolor{black}{lossy compression can be applied to large data structures like activations or caches to mitigate memory bottlenecks\cite{han2016deep_compression,jacob2018quantization, banner2019post,frantar2022gptq, jin2019deepsz, jin2021comet}}. 
For example, quantization is a form of lossy compression that reduces the precision of numerical data by mapping continuous or high-precision values (e.g., float32) to a smaller, discrete set of values such as integers.
% A prominent technique for lossy compression in this domain is quantization\cred{xh: some citations are needed here}. Quantization reduces the precision of numerical data, mapping continuous or high-precision values (e.g., floating-point numbers like float16) to a smaller, finite set of discrete values (often integers). 
In fact, this inherently reduces the entropy of the data, making it smaller and often more amenable to entropy encoding.
\textcolor{black}{In image, video, and scientific lossy compression, entropy encoding and spatial encoding are commonly added to further reduce data size\cite{wallace1992jpeg,taubman2002embedded,lindstrom2014fixed,di2016fast, ohm2012comparison, di2016fast, tao2017significantly, sz18}}
However, the low throughput of these post-quantization encoders makes them impractical for real-world LLM inference, where latency is critical. 
For example, \textcolor{black}{CacheGen\cite{liu2024cachegen} integrates such encoders into the quantization pipeline and achieves throughput below 1 GB/s, which is an improvement over network transmission, but insufficient compared to the performance of GPU-CPU memory migration.}
% This inherently reduces the entropy of the data, making it smaller and often more amenable to subsequent lossless compression stages. \cred{[we need to introduce cachegen here]}\cb{bo: because we have not implemented offloading system, there is no need to mention infinigen}
% In this paper, we propose a GPU-based high-speed entropy encoder that enables post-quantization encoding for LLM inference, significantly improving KV cache compression ratios with minimum overhead.

Another challenge in applying lossy compression is to manage the trade-off between the desired compression ratio and the potential impact on model accuracy.
% The core challenge in applying lossy compression, especially quantization, lies in managing the trade-off between the desired compression ratio and the potential impact on the model's performance or accuracy. 
Applying quantization requires careful consideration of the data's characteristics and the specific downstream task. Different quantization strategies, such as varying granularity (e.g., token-wise, channel-wise, block-wise) or using adaptive error bounds, can be employed to control the information loss and minimize accuracy degradation~\cite{liu2024kivi}. 

\subsection{GPU architecture}

% Graphics Processing Units (GPUs) have undergone a remarkable evolution from their origins as specialized hardware for rendering graphics to becoming powerful general-purpose computing engines. This transformation has been particularly significant for deep learning applications, where GPUs now serve as the backbone for large language model (LLM) inference workloads.

% GPU is designed with parallelism through thousands of cores organized into Streaming Multiprocessors (SMs). 
% This design is well-suited for LLM operations, where large-scale matrix computations dominate. 
% GPU cores execute kernels using a large number of concurrent threads, typically organized into thread blocks to optimize scheduling and data sharing.

% The fundamental architecture of GPUs differs substantially from traditional CPUs, being designed around a paradigm of massive parallelism. While CPUs excel at sequential processing with sophisticated control logic and large caches, GPUs thrive when executing the same operation across vast amounts of data simultaneously—precisely the pattern found in many LLM operations. At the heart of GPU architecture is its parallel processing capability, with thousands of cores organized into Streaming Multiprocessors (SMs). These cores execute kernels through a multitude of concurrent threads, often grouped into thread blocks for organizational efficiency.

\textcolor{black}{The GPU memory hierarchy plays a crucial role in performance optimization\cite{mei2016dissecting}.}
Global memory provides the largest capacity with high latency, making efficient access patterns essential. 
\textcolor{black}{For example, coalesced memory access can significantly improve bandwidth utilization.}
While shared memory in each SM is a programmer-managed cache shared among threads within a block.
% enabling fast data exchange and reuse. 
% Registers offer thread-private storage for immediate computations, while hardware-managed L1 and L2 caches help bridge the latency gap for less predictable memory accesses.

% The memory hierarchy of GPUs plays a crucial role in performance optimization. Global memory offers the largest capacity but comes with higher latency, making access patterns critical—coalesced memory access, where threads read contiguous memory locations, significantly improves bandwidth utilization. Closer to the computation units, shared memory serves as a programmer-controlled cache shared among threads in a block, enabling efficient data exchange and reuse. At the fastest level, registers provide thread-private storage for immediate computations, while hardware-managed L1 and L2 caches help bridge the latency gap for memory accesses.

GPU execution follows a Single Instruction, Multiple Thread (SIMT) model, where threads within a warp (typically 32 threads) execute the same instruction in lockstep. This model creates challenges when threads need to follow different execution paths, which is also called branch divergence that forces serialization and reduces performance. 
To coordination among threads, GPUs provide atomic operations that ensure read-modify-write sequences complete without interference, essential for managing shared resources or global indices across thread blocks.

% Effective GPU optimization for LLM inference requires a deep understanding of these architectural elements. By aligning algorithms with the GPU's parallel nature and memory hierarchy, developers can achieve the high throughput necessary for efficient LLM inference, particularly when implementing sophisticated compression and attention mechanisms.
% \input{tex/03_Observation}
\section{Design Methodology}
\label{sec:4.0}

\begin{figure*}[tbp]
    \centering
    \begin{subfigure}[]{0.16\textwidth}
        \centering
        \includegraphics[width=\textwidth]{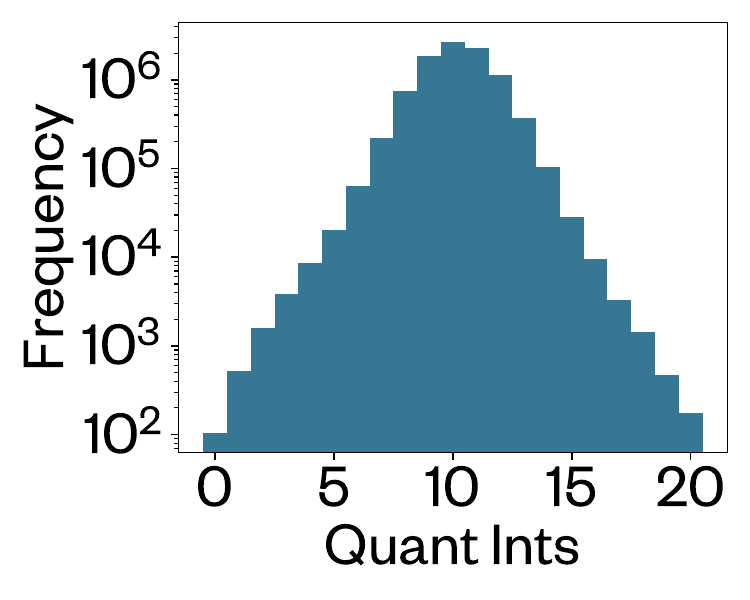}
        \vspace{-0.7cm}
        \caption{K-0}
        \label{fig:hist_k_1}
    \end{subfigure}
    \hfill
    \begin{subfigure}[]{0.16\textwidth}
        \centering
        \includegraphics[width=\textwidth]{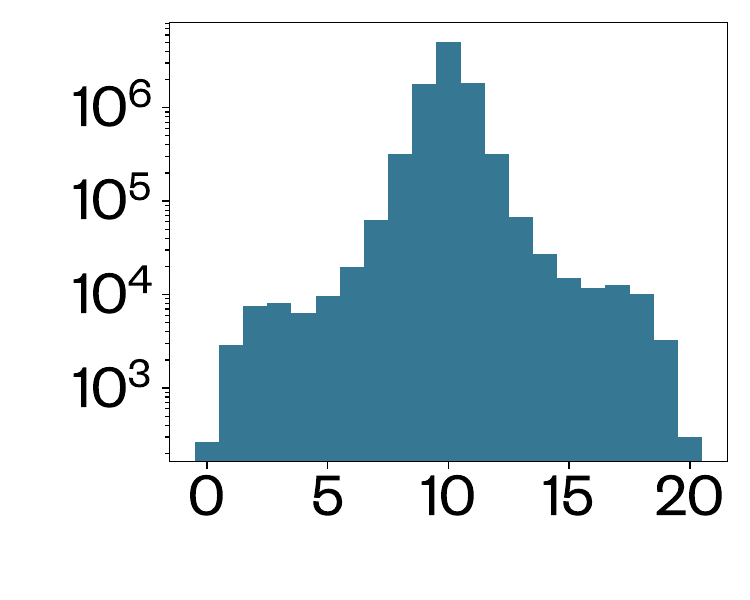}
        \vspace{-0.7cm}
        \caption{K-19}
        \label{fig:hist_k_20}
    \end{subfigure}
    \hfill
    \begin{subfigure}[]{0.16\textwidth}
        \centering
        \includegraphics[width=\textwidth]{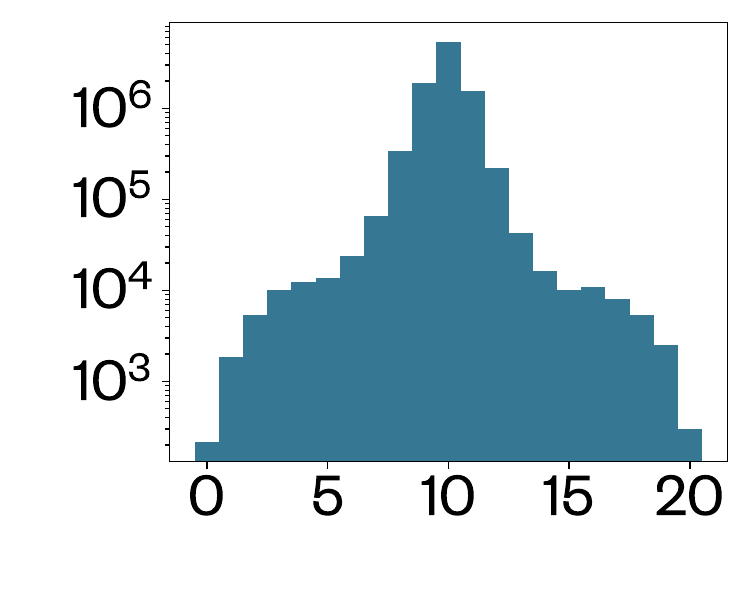}
        \vspace{-0.7cm}
        \caption{K-39}
        \label{fig:hist_k_39}
    \end{subfigure}
    \hfill
    \begin{subfigure}[]{0.16\textwidth}
        \centering
        \includegraphics[width=\textwidth]{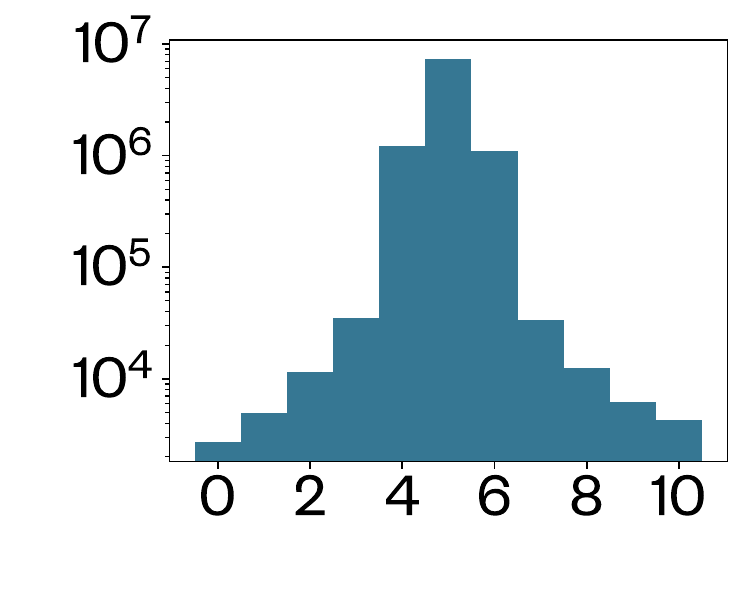}
        \vspace{-0.7cm}
        \caption{V-0}
        \label{fig:hist_v_1}
    \end{subfigure}
    \hfill
    \begin{subfigure}[]{0.16\textwidth}
        \centering
        \includegraphics[width=\textwidth]{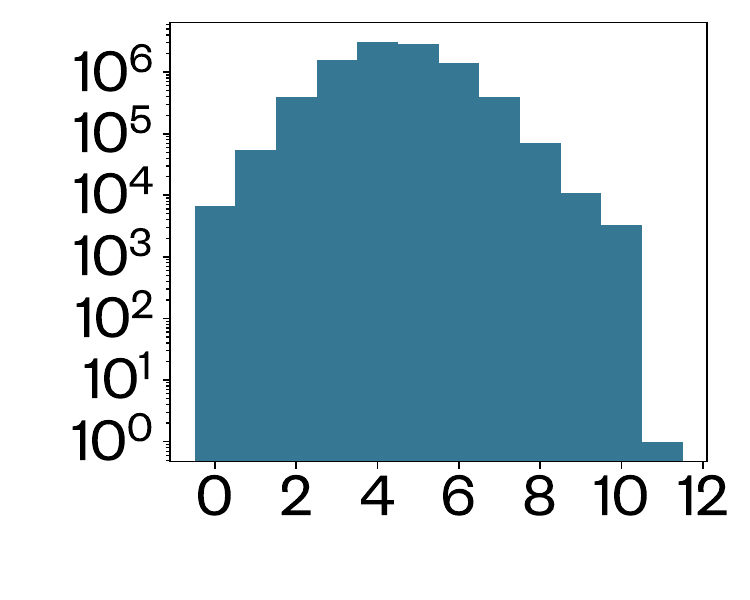}
        \vspace{-0.7cm}
        \caption{V-19}
        \label{fig:hist_v_20}
    \end{subfigure}
    \hfill
    \begin{subfigure}[]{0.16\textwidth}
        \centering
        \includegraphics[width=\textwidth]{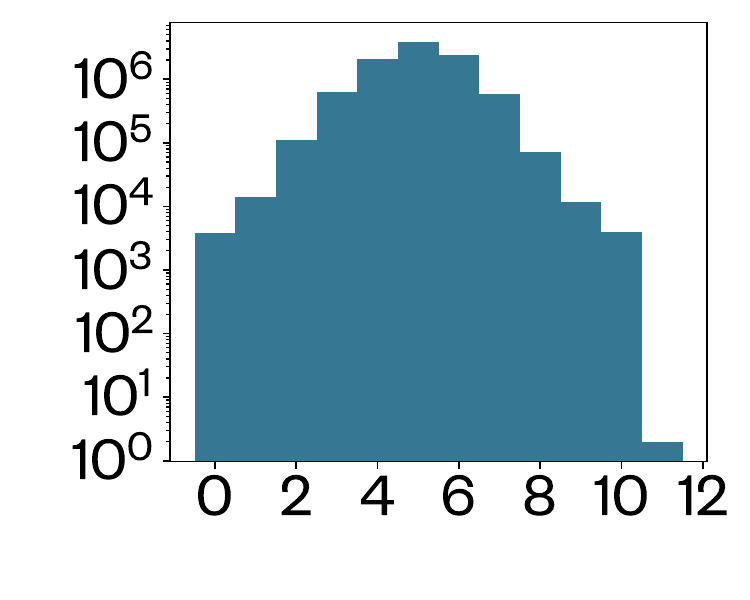}
        \vspace{-0.7cm}
        \caption{V-39}
        \label{fig:hist_v_39}
    \end{subfigure}
    % \vspace{-0.3cm}
    \caption{Histogram of quantized KV cache based on GSM8k benchmarking of Llama2-13b. K values are block-wise quantized, while V values are token-wise quantized.}
    \Description{Six histogram plots arranged in two rows showing frequency distributions of quantized values. Top row shows K values (K-0, K-19, K-39) with different layer indices, bottom row shows V values (V-0, V-19, V-39) demonstrating various quantization patterns across different transformer layers.}
    \label{fig:all_histograms}
    \vspace{-0.3cm}
\end{figure*}

This section presents our lossy KV cache compression framework designed for large language model generative inference. 
% [we need a figure here to show the architecture of the KV cache lossy compressor]. 
Figure~\ref{fig:overview} presents an overview of the proposed framework, which is organized into two stages: \blackcircled{1}-\blackcircled{4} (\textbf{Store}) and \redcircled{1}-\redcircled{4} (\textbf{Fetch}).

During the \textbf{Store} stage, including the LLM inference prefill phase, the KV cache generated from processing the user prompt is immediately compressed in a 2D blockwise manner to significantly reducing its memory footprint.
% During the \textbf{Store} stage, for the LLM inference prefill phase, the KV cache generated from processing the user prompt is immediately compressed in a 2D blockwise manner, significantly reducing its memory footprint. 
During the subsequent decode phase, newly generated KV cache vectors are accumulated in a fixed-size buffer. Once the buffer overflows, the compressor segments the buffer into 2D blocks, compresses these blocks, and appends the compressed data to the previously stored KV cache as described in Section~\ref{data_appending}.
Specifically, \blackcircled{1}, once the buffer exceeds its optimal size, we segment its contents in place into integer multiples of the block size. For each 2D block, we assign a GPU thread block to efficiently load the data into shared memory and perform quantization in situ. \blackcircled{2}, based on the requirements of matrix-vector multiplication, each thread encodes a slice of the 2D block residing in shared memory, storing the resulting bits in the thread’s buffer. \blackcircled{3}, threads aggregate these encoded data within shared memory. \blackcircled{4}, all threads cooperatively write the compressed data back to a contiguous region in GPU global memory. To prevent write conflicts, a global atomic variable governs the write-back offset. Furthermore, the compression thread block records this offset in a \textit{Block Offsets Array} to facilitate subsequent retrieval during the \textbf{Fetch} stage.

We propose a novel quantization–entropy-encoding pipeline for compressing the KV cache. 
Specifically, we preallocate a dedicated buffer for each layer’s KV cache to maintain newly generated cache vectors until the buffer reaches suitable size for compression. 
Then, we take advantage of the statistical properties of the KV cache by carefully selecting quantization units and applying quantization with a fixed relative error bound, which translates into an absolute error bound for each unit. 
Next, we generate the histogram of quantization codes on the GPU and construct shared Huffman codebooks on the CPU. 
Note that these codebooks are generated once during the prefill stage and reused during runtime, avoiding repeated overhead. 
Finally, we encode the quantized KV cache using our high-throughput, GPU-optimized entropy encoder.

During the \textbf{Fetch} stage, our framework adopts a just-in-time approach to minimize decompression overhead during inference.
% During the \textbf{Fetch} stage, to minimize decompression overhead during inference, our framework adopts a just-in-time approach. 
% When the model requires access to the KV cache, each block is independently decompressed directly into GPU shared memory and registers. 
% This enables the model to immediately use the decompressed data for matrix multiplications without incurring additional memory transfers.
It is important to note that data flow during the \textbf{Fetch} stage is significantly more intensive than during the \textbf{Store} stage: each KV vector is compressed and stored only once but is fetched multiple times for newly generated token. 
Thus the decompression and memory access pattern dominate the overall performance impact, making them the primary targets of optimization in our design.
Specifically, \redcircled{1}, we launch GPU thread blocks independently for each 2D block, loading compressed data into shared memory using the \textit{Block Offsets Array}. 
\redcircled{2}, each thread decompresses a slice of data, mirroring the assignment from the \textbf{Store} stage. \redcircled{3}, dequantization is performed, and the resulting data is immediately used for dot product computations in GPU registers and shared memory as appropriate. This design minimizes global memory accesses and accelerates both decompression and attention operations. 
\redcircled{4}, shared memory serves as a buffer to write the results of matrix-vector multiplications efficiently.

\subsection{Compression Overview}
\label{sec3.1}

Our compression pipeline comprises two main components: \textbf{Lossy Compression via Quantization:} This step reduces data entropy by mapping high-precision values to a finite set of discrete integers. \textbf{Lossless Compression via Entropy Encoding:} the quantization codes are encoded using high-throughput GPU-optimized encoding.

\subsubsection{Quantization}

Quantization is the only lossy step in our compression pipeline that may impacts model accuracy. 
% It is the only step that impacts model accuracy. 
Consequently, our method achieves same accuracy to that of standalone quantization while providing additional memory savings via subsequence entropy encoding.
We observe that several quantization granularities are can be applied to the KV cache tensor [context\_len, head\_num, head\_dim]. 
Inspired by KIVI~\cite{liu2024kivi}, we partition the key (K) cache along the \texttt{context\_len} dimension into fixed-size blocks and apply channel-wise quantization within each block. For the value (V) cache, we adopt a token-wise quantization strategy.

% revise_01
% Given that the KV cache tensor has the shape \texttt{[context\_len, head\_num, head\_dim]}. There are only a limited number of quantization granularity available. Building on the findings from \cred{KIVI—the state-of-the-art method}. for KV cache quantization—we first partition the key (K) cache along the context length dimension into fixed-size blocks and then apply channel-wise quantization within each block. For the value (V) cache, we adopt a token-wise quantization strategy. Our quantization will be based on KIVI's findings.

\begin{figure*}[tbp]
    \centering
   \includegraphics[width=0.9\linewidth]{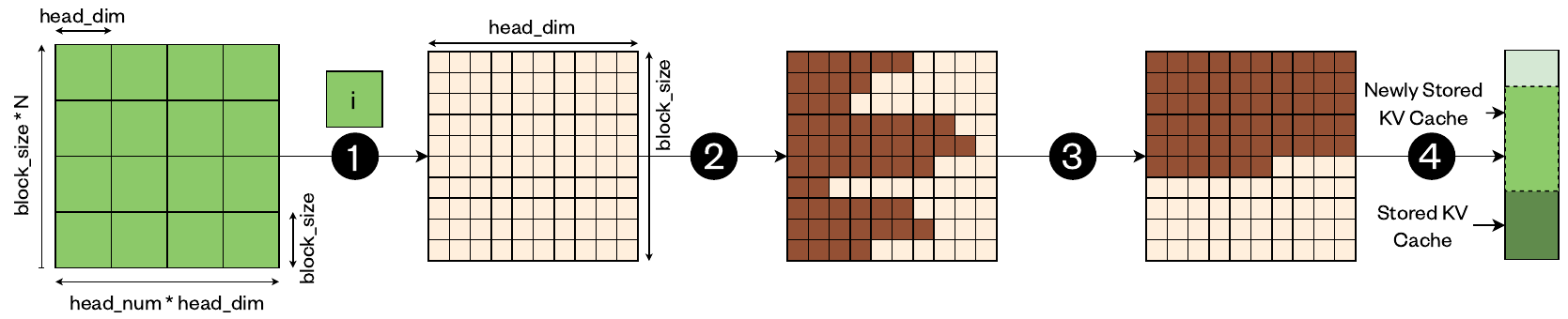}
   % \vspace{-0.2cm}
    \caption{Compression Overview. 1. For each 2D block in the KV cache, assign a unique index that indicates the block's position within the entire cache. 2. Each 2D block is loaded into shared memory in a coalesced manner. 3. According to the dot product direction, each vector is encoded into bits and stored in the thread's shared memory buffer. 4. An in-block inclusive scan is performed to compute bit offsets for every thread, aggregating all bits into a contiguous chunk in shared memory. 5. The atomic offset is updated, and all threads coalescedly write the chunk back to global memory.}
    \Description{Step-by-step compression pipeline diagram showing 5 stages: 2D block indexing, coalesced loading to shared memory, vector encoding with bit buffers, inclusive scan for bit aggregation, and atomic offset updates for global memory writes.}
    \label{fig:compress}
    % \vspace{-0.2cm}
\end{figure*}

\subsubsection{Entropy Encoding}

Following quantization, the histogram of integers reveals that the distribution of the quantized values is highly skewed, with most values concentrated around a few discrete levels, shown in figure~\ref{fig:all_histograms}. This redundancy makes Entropy Encoding an ideal option to further reduce data size. We propose to adopt Huffman-based encoding to encode the quantization codes. Compared to alternatives such as arithmetic encoding\cite{witten1987arithmetic} or Asymmetric Numeral Systems (ANS)\cite{duda2009asymmetric}, Huffman-based encoding shows great balance between simplicity and efficiency for KV cache encoding based on our evaluation.

% revise_02
\subsubsection{Compression Design Trade-offs.}
% Considering data characteristics, we can use quant + lorenzo + entropy encoding. But we have no theory to
%  explain why this combination works well and only this combination works well.
Based on our experiments, we acknowledge the effectiveness of adding a Lorenzo predictor before quantization to capture spatial information. 
However, to meet the performance requirements of large language model inference, we prioritize cache-resident decompression to accelerate the decoding process. 
Moreover, the computational patterns of the Lorenzo predictor and Huffman encoding direction conflict with each other. 
For K cache, the dot product computation in matrix-vector multiplication occurs along the head\_dim dimension, whereas the inverse operation of the Lorenzo predictor follows a sequential dependency in the vertical direction during decompression. 
% This fundamental mismatch in data access patterns makes it difficult to efficiently integrate the two operations.
Integrating pre-quantization predictors, entropy decoding, and matrix-vector multiplication into a unified pipeline requires allocating substantial shared memory per GPU thread block to serve as an intermediate buffer, which introduces significant overhead to the decompression process. 
As a result, we conclude that incorporating such predictors before quantization result in a computational cost that outweighs the marginal compression gains.
By decompressing data directly into cache and feeding it immediately to the matrix-vector multiplication kernels, we eliminate the need to write decompressed data back to global memory and reduce memory access overhead where compressed data is significantly smaller than original.

\subsection{Compression Design}

In the prefill stage, the LLM processes a user prompt containing thousands of tokens, with each token generating a fixed-dimension vector (shape \texttt{[head\_num,}\texttt{ head\_dim]}) per transformer layer. Across all layers, this results in \texttt{layer\_num $\times$ 2} vectors (two per layer corresponding to the key and value caches). Due to statistical differences between layers, we generate shared Huffman codebooks for each layer during LLM inference prefill stage. 
% revise_03

\subsubsection{Buffering and Blocking}

For both K and V cache, we maintain a buffer of shape \texttt{[buffer\_ctx\_len, head\_num, head\_dim]}, where \texttt{buffer\_ctx\_len} ranges from 1 to \texttt{buffer\_size}. When the buffer overflows, it is truncated such that its size falls within the allowed range and can be partitioned into complete blocks. The overview of compression design shwon in the figure~\ref{fig:compress}.

% revise_05
% [we would need a figure here to illustrate the buffer truncation and blocking process]

The K cache is considered as a tensor with shape [ \text{context\_len}, \(\text{head\_num} \times \text{head\_dim}\)].
We map it to 2D blocks for compression.
Since the matrix-vector multiplication dot product operates along the \(\text{head\_dim}\) dimension, we must use \(\text{head\_dim}\) as one of the dimensions of the 2D blocks when assign the buffer into complete blocks. 
The \(\text{context\_len}\) is used as the other dimension of the 2D blocks, and the tensor is split into \(\text{head\_num}\) parts.
% is variable and depends on the user setting and GPU shared memory capacity.
% The \(\text{head\_num} \times \text{head\_dim}\) dimension is split into \(\text{head\_num}\) parts. 
% For the other edge of the 2D block, we use the \(\text{context\_len}\) dimension, which is variable and depends on the user setting and GPU shared memory capacity.

For the V cache, the vector dot product in matrix-vector multiplication occurs along the \(\text{context\_len}\) dimension, we divide the tensor to \text{head\_dim}\ blocks (i.e., the same as K cache) to simplify the kernel implementation for the compression of both K and V caches.

\subsubsection{GPU-based Entropy Encoding}
\label{sec:huffman_encoding}

Once blocked, the compression process is executed on the GPU and consists of the following six stages:
\textbf{Coalesced Memory Loading:} The 2D block is loaded into shared memory.
\textbf{Huffman-based Encoding:} Each thread compresses its assigned data slice of a row or column of the 2D data block in shared memory. Usually a head\_dim length vector per thread.
\textbf{Inclusive Scan:} An in-block inclusive scan computes bit offsets.
\textbf{Data Aggregation:} Encoded data is aggregated in shared memory for coalesced global memory write-back.
\textbf{Index Synchronization:} A wait-free, atomic operation synchronizes the global memory write-back index.
\textbf{Coalesced Global Memory Write-back:} Compressed data is written back to global memory in a coalesced manner.

As shown in Figure~\ref{fig:compress}, in stage \blackcircled{1} of the compression process, we assign a unique index to each 2D block in the overflowed, truncated KV cache chunk. This index identifies both the attention head to which the 2D block belongs and its corresponding \texttt{ctx\_len} range. For each 2D block, we launch a GPU thread block containing \texttt{block\_size} threads to perform compression. For K cache compression, threads within a block access contiguous memory addresses in the 2D block of shape \([block\_size, head\_dim]\), loading the data into shared memory. Within each GPU block, every thread is responsible for compressing one specific row of the K cache, each row having a shape of \([head\_dim]\). Similarly, for each V cache 2D block with shape \([head\_dim, block\_size]\), each thread processes a specific column, matching the shape of the K cache. In step \blackcircled{2}, each thread then encodes one row or column of data in shared memory. To minimize bank conflicts, we set the shape of the shared memory 2D block to ensure that both K and V compressions can access shared memory efficiently. Prior to the encoding process, the shared Huffman codebooks is loaded into shared memory to enable efficient encoding.

Once all threads have finished encoding, in step \blackcircled{3}, the bit counts produced during entropy encoding are accumulated. Each thread within the block performs an inclusive scan using the CUB library to determine its bit offset within the aggregate buffer. After obtaining the total bit count, each thread writes its compressed bits to a shared memory buffer for aggregation. In the final step \blackcircled{4}, the block performs a global atomic operation that updates a global variable to acquire the latest available index for writing the compressed data to global memory. This is followed by a coalesced write of the compressed data to GPU global memory. The atomic operation is crucial for preventing write-back conflicts between GPU thread blocks—without it, two thread blocks might retrieve and update the global offset variable simultaneously, leading to data being written to the same address and resulting in decompression errors.

Our 2D blockwise compression design is highly flexible and readily integrates with mainstream inference frameworks \textcolor{black}{such as vLLM\cite{kwon2023efficient} and SGLang\cite{zheng2024sglang}}, both of which employ block-wise KV cache management strategies. 
% Owing to our parallel compression design, the 
Additionally, our framework can easily extend to other encoding algorithms, which can efficiently process \texttt{head\_dim} unsigned 8-bit integers. For example, our framework can be adapted to encoding methods such as FSE\cite{collet2013fse}, Bitshuffle\cite{masui2015compression, zhang2023fz}, and fixed-length encoding\cite{huang2023cuszp}.

Although our parallel approach enables fine-grained Huffman encoding, it introduces memory overhead from additional metadata. Specifically, each thread needs an unsigned 16-bit integer to store the number of bits generated during encoding, while each block requires an unsigned 32-bit integer to store the global memory index for the compressed data. Given the average number of bits per encoded unsigned 8-bit integer is approximately 2, the total memory overhead for thread metadata is \(16 / (\texttt{head\_dim} \times \texttt{avg\_bits\_per\_integer})\), which corresponds to approximately \(\frac{1}{16}\) of the compressed data size, or \(\frac{1}{128}\) of the original data size. The per-block 32-bit integer used for the global memory index is even less significant, contributing less than \(\frac{1}{16}\) of the compressed data size. Therefore, the overall memory overhead for metadata is negligible compared to the size of the compressed data.

% \subsection{Decode Stage}

\subsubsection{Natural Data Appending}
\label{data_appending}

During the decoding stage of LLM inference, each decoding step produces a new cache vector of shape [\texttt{head\_num}, \texttt{head\_dim}], with the entire KV cache for a given layer participating in the attention computation.
We propose to preallocate a buffer that can accommodate up to \texttt{buffer\_size} newly generated cache vectors. The buffer is truncated only when the number of cache vectors exceeds \texttt{buffer\_size}, ensuring efficient and consistent data management.

Each newly generated 2D block receives a unique block index that continues from the last used index in the compression sequence. These new cache vectors are then compressed using our encoding kernel. The compressed data for each 2D block remains independent due to our blockwise compression design, allowing us to seamlessly append the newly compressed data to the end of the existing compressed data buffer to allow efficient storage and effective management of the compressed cache over time.

In practice, repeated kernel launches on GPUs can introduce substantial performance overhead and limit maximum throughput due to high launch latency. 
For example, in LLaMA2-13B inference with a context length of 32K, block size of 64, and 512 separate kernel launches, the cumulative launch time limits throughput to approximately 61 GB/s. 
To address this, we employ a single kernel launch with multiple parallel GPU blocks to decode compressed data concurrently, effectively minimizing the overhead associated with frequent kernel invocations during decoding.

% % For example, when the context length reaches 32K tokens---a common scenario in current large language model applications---and the \texttt{block\_size} is set to 64, the absence of contiguous data appending requires launching approximately 512 separate kernels to fetch the data. Each kernel launch typically incurs about 15 microseconds of overhead, resulting in a total launch overhead of approximately $512 \times 15\,\mu\mathrm{s} = 7.68\,\mathrm{ms}$. As a result, the cumulative kernel launch latency becomes the main bottleneck in the retrieval process.

% % Taking Llama2-13B as an example, the key or value cache for a single layer with a context length of 32K can be computed as:
% \(
% 40~(\text{head\_num}) \times 128~(\text{head\_dim}) \times 32{,}768~(\text{sequence length}) \times 2~\mathrm{B}~(\text{FP16}) = 320\,\mathrm{MB}.
% \)
% Even without accounting for any other computational overhead, this kernel launch time alone would limit the throughput to approximately $42\,\mathrm{GB/s}$ (since $320\,\mathrm{MB} / 7.68\,\mathrm{ms} \approx 41.7\,\mathrm{GB/s}$), making it the primary bottleneck under such settings.

\subsection{Decompression Design}
\label{sec:decompression}

During the decode stage of LLM inference, each iteration compress the KV cache of only one input token while requiring decompression of the entire contextual KV cache. This creates a significant \textbf{imbalance} in the demand for compression versus decompression operations. Therefore, in our evaluation and performance optimization, we focus primarily on decompression throughput.

\subsubsection{High-throughput GPU Decoding.}
One of the major bottlenecks in decompression is Huffman-based decoding coupled with the global memory write-back of decompressed data. 
To mitigate this, we redesign the decoding process to improve the decompression efficiency. 
One of the classic GPU Huffman decoding implementations is from cuSZ\cite{tian2020cusz}. However, cuSZ adopts a coarse-grained block partitioning strategy that limits the achievable parallelism during decompression. 
Moreover, its bitwise decoding involves frequent if-else branching within the decoding loop for each bit, resulting in excessive thread divergence and underutilization of GPU resources. 
% These inefficiencies lead to suboptimal performance in highly parallel decoding scenarios, making the cuSZ approach unsuitable for the requirements of our project. 
In comparison, our solution adopt fine-grained parallel encoding that achieves around 256 bits of decoding per thread, which can even reduce to 128 bits considering that the meta data overhead is negligible.
% In Section~\ref{sec:huffman_encoding}, we introduced a GPU-based, fine-grained parallel Huffman encoding method that achieves around 256 bits of decoding per thread which can even reduce to 128 bits considering that the meta data overhead is negligible. 
% This approach further highlights the challenge of branch divergence, which we address in detail in the following section to enable higher throughput for KV cache decompression tasks. 
% Besides, due to our cache-resident decompress design, to achieve minimum data transfer of the decompress and matrix vector multiplication process, we need to assign each threads 

% revise_04

% \begin{algorithm}
% \caption{Huffman Decoding Algorithm}
% \SetAlgoLined
% $node\_ref \gets root\_node$\;
% $write\_idx \gets 0$\;
% \While{bit available in bit stream}{
%     \If{$node\_ref.is\_symbol = true$}{
%         $decode\_buffer[write\_idx++] = node\_ref.symbol$\;
%         $node\_ref \gets root\_node$\;
%     }
%     Read next bit from bit stream\;
%     \eIf{bit = 1}{
%         $node\_ref \gets node\_ref.one\_branch$\;
%     }{
%         $node\_ref \gets node\_ref.zero\_branch$\;
%     }
% }
% \end{algorithm}

% \cred{xh: we need to explain this issue before we address it.} \cb{bo: add the information in the last paragraph}

To address the branch divergence issue due to the fact that all threads within a warp execute the same instruction simultaneously on GPUs,
% of the original cuSZ Huffman decoding kernel, 
we redesign the decoding process to improve its efficiency. 
% In GPU architectures, all threads within a warp execute the same instruction simultaneously. Therefore, to optimize the efficiency of Huffman decoding on GPUs, we must design a branch divergence-free procedure that can be executed in parallel by all threads in a warp. 
% Back to the concept view of Huffman decoding, the Huffman decoding process can be viewed as a tree traversal. 
We consider Huffman-based decoding as a tree traversal.
Starting at the root of the tree, we update a pointer instead of using explicit conditional statements to navigate toward the leaf nodes.
Specifically, we propose two major optimizations:

% At each node, we check whether the current bit is 0 or 1. 
% If it is 0, we move to the left child; if it is 1, we move to the right child. 
% This traversal continues until a leaf node is reached, at which point we have decoded a symbol. In this tree traversal, the key operation is checking whether the current bit is 0 or 1, and updating the pointer to move left or right accordingly. This process is the shared procedure that can be executed in parallel by all the threads within a warp. 

\textbf{Array-Based Representation of the Huffman Tree:} Due to the frequent access requirements of the Huffman decode tree, we must store the codebook in GPU shared memory to achieve high-speed access, which requires a compact representation of the Huffman tree. To efficiently store a binary tree in contiguous memory, we arrange the Huffman tree nodes adjacent to one another. In this representation, traditional pointers in each node are replaced by indexes that reference the node's position in the node array. Each parent node contains two indexes representing references to its child nodes. 

\textbf{Branch Divergence-Free Approach:} 
Rather than using explicit conditional statements to navigate the Huffman tree, we employ branchless operations for decoding to minimize branch divergence in parallel environments. Specifically, during tree traversal, each internal node stores its child node indexes in a two-element array, and the bit value from the bit stream is used directly as the index to select the next node—if the bit is 0, traversal proceeds to \texttt{indexes[0]}; if it is 1, to \texttt{indexes[1]}. For symbol handling, two key optimizations are applied: First, at each iteration, the symbol from the current node is always written to the decode buffer, but the buffer's write position is only advanced when the node represents a symbol. This is achieved by updating the write index using \(write\_idx += nodes[index].is\_symbol\), so that only symbol nodes increment the output position. Second, returning to the root node after decoding a symbol is implemented with bitwise logic: the tree position is reset by computing \(index \&= \sim(-nodes[index].is\_symbol)\), which sets \texttt{index} to 0 if the current node is a symbol node (\(is\_symbol = 1\)), and leaves \texttt{index} unchanged otherwise (\(is\_symbol = 0\)). 
By avoiding all conditional branches inside the decoding loop, our approach achieves substantial reductions in branch divergence and improve efficiency for parallel decoders. 
The decoding process iterates over the bit stream, using each bit to select the next node in the tree and applying these branchless updates, until all bits have been processed.

% \begin{algorithm}
% \caption{Branch Free Huffman Decoding}
% \label{alg:branch_free_array_huffman_decode_optimized}
% \SetAlgoLined
% $index \gets 0$\;
% $write\_idx \gets 0$\;
% \While{bit available in bit stream}{
%     $decode\_buffer[write\_idx] = nodes[index].symbol$\;
%     $write\_idx += nodes[index].is\_symbol$\;
%     $index\&=\sim(-nodes[index].is\_symbol)$\;
%     Read next bit from bit stream\;
%     $index = nodes[index].indexes[bit]$\;
% }
% \end{algorithm}

% \begin{algorithm}
% \caption{Branch Free Huffman Decoding}
% \label{alg:branch_free_array_huffman_decode_optimized}
% \KwIn{
%     $bit\_stream$: The encoded bit stream to be decoded\;
%     $nodes$: Array representation of the Huffman tree, where each node contains the fields \texttt{is\_symbol}, \texttt{symbol}, and \texttt{indexes[2]}\;
% }
% \KwOut{
%     $decode\_buffer$: The decoded sequence of symbols\;
% }
% $index \gets 0$\;
% $write\_idx \gets 0$\;
% \While{bit available in bit stream}{
%     $decode\_buffer[write\_idx] = nodes[index].symbol$\;
%     $write\_idx += nodes[index].is\_symbol$\;
%     $index\ \&= \sim(-nodes[index].is\_symbol)$\;
%     Read next bit from bit stream\;
%     $index = nodes[index].indexes[bit]$\;
% }
% \end{algorithm}

\begin{figure*}[tbp]
    \centering
    \begin{subfigure}[]{0.33\textwidth}
        \centering
        \includegraphics[width=\textwidth]{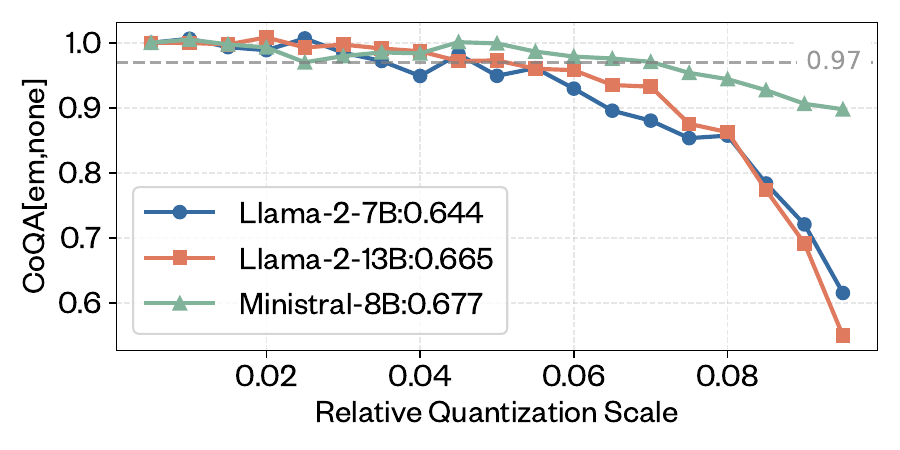}
        \caption{K BlockQuant}
        % \label{fig:k_block_q}
    \end{subfigure}
    \hfill
    \begin{subfigure}[]{0.33\textwidth}
        \centering
        \includegraphics[width=\textwidth]{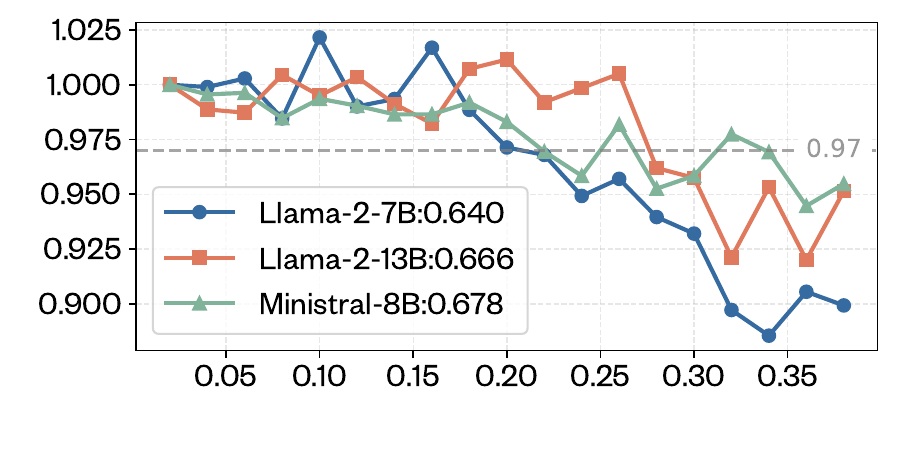}
        \caption{K ChannelQuant}
        % \label{fig:hist_k_39}
    \end{subfigure}
    \hfill
    \begin{subfigure}[]{0.33\textwidth}
        \centering
        \includegraphics[width=\textwidth]{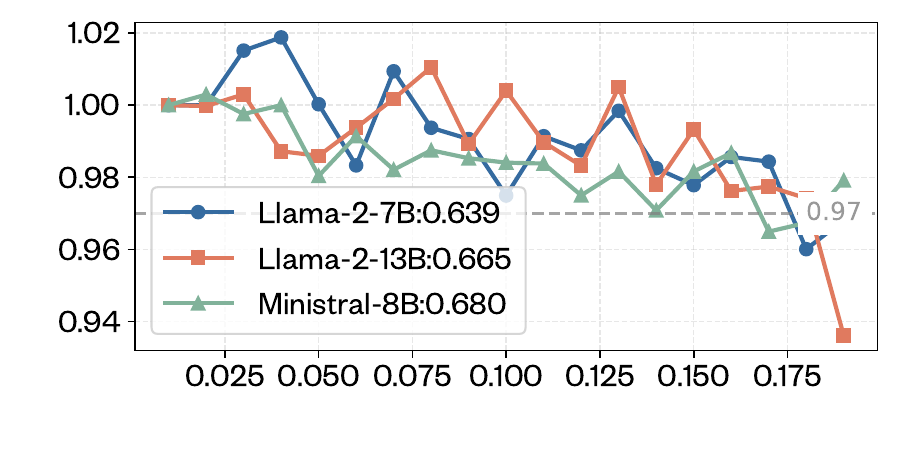}
        \caption{V TokenQuant}
        % \label{fig:hist_k_39}
    \end{subfigure} 
    \vspace{-0.2cm}
    \caption{Accuracy(Nomalized, CoQA[em,none]) vs KV standalone Relative quantization scale from three models. The numbers to the right of the legend show the value that corresponds to the normalized value of 1 for each model.} 
    \Description{Three line plots showing accuracy vs quantization scale for K BlockQuant, K ChannelQuant, and V TokenQuant methods. Each plot shows declining accuracy curves for three different language models as quantization scale increases, with accuracy normalized to 1.0 at baseline.}
    % \vspace{-0.3cm}
    \label{fig:k_v_separate}
\end{figure*}

\begin{figure*}[t]
    \centering
    
    \begin{subfigure}[b]{0.24\textwidth}
        \centering
        \includegraphics[width=\textwidth]{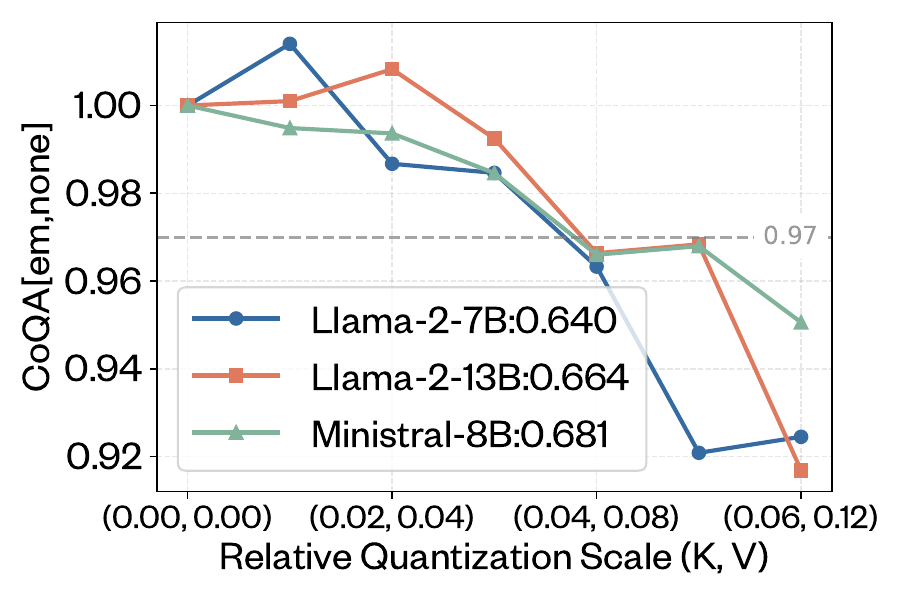}
        \caption{K BlockQuant, CoQA}
        % \label{fig:k_block_q}
    \end{subfigure}
    % \hfill
    \begin{subfigure}[b]{0.24\textwidth}
        \centering
        \includegraphics[width=\textwidth]{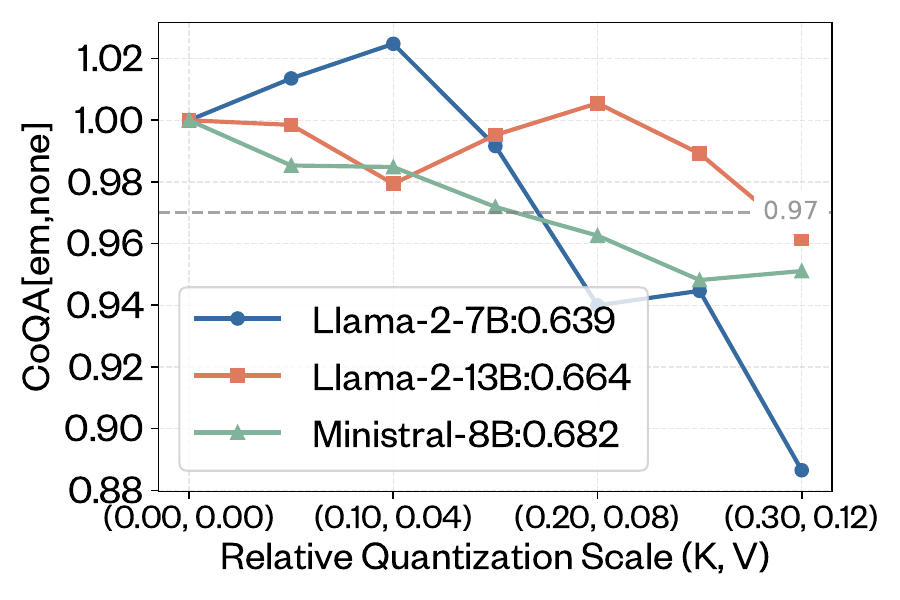}
        \caption{K ChannelQuant, CoQA}
        % \label{fig:k_block_q}
    \end{subfigure}
    % \vspace{1em}
    \begin{subfigure}[b]{0.24\textwidth}
        \centering
        \includegraphics[width=\textwidth]{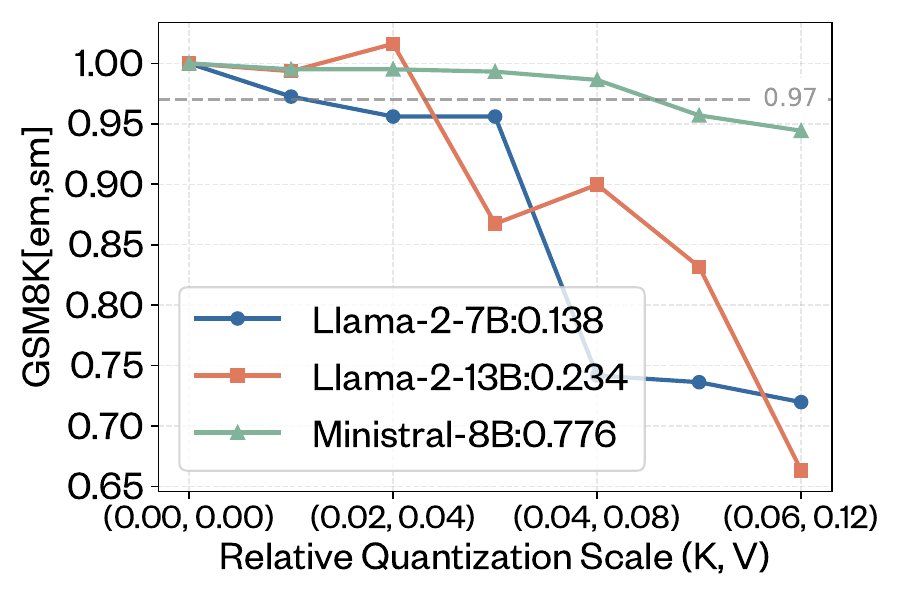}
        \caption{K BlockQuant, GSM8K}
        % \label{fig:k_block_q}
    \end{subfigure}
    % \hfill
    \begin{subfigure}[b]{0.24\textwidth}
        \centering
        \includegraphics[width=\textwidth]{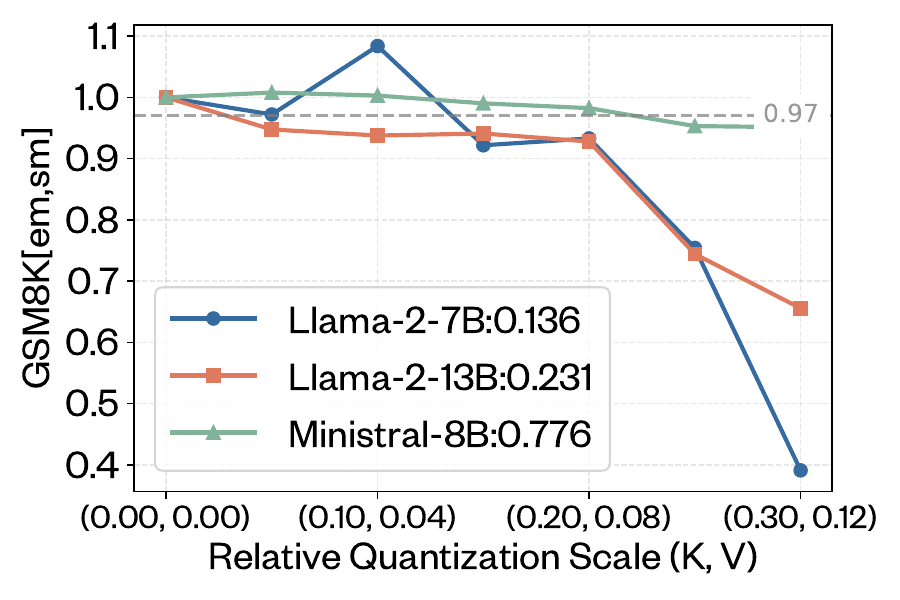}
        \caption{K ChannelQuant, GSM8K}
        % \label{fig:k_block_q}
    \end{subfigure}
    \vspace{-0.2cm}
    \caption{Accuracy(Nomalized, CoQA[em,none], GSM8K[em,sm]) vs Relative quantization scale([K,V]) from three models. The numbers to the right of the legend show the value that corresponds to the normalized value of 1 for each model.}
    \Description{Four heatmap plots showing accuracy vs combined K-V quantization scales. Two plots for CoQA (BlockQuant and ChannelQuant) and two for GSM8K, displaying how accuracy degrades with different combinations of K and V compression levels across different models.}
    \label{fig:k_v_combine}
\end{figure*}

\begin{figure*}[t]
    \centering
    
    \begin{subfigure}[b]{0.24\textwidth}
        \centering
        \includegraphics[width=\textwidth]{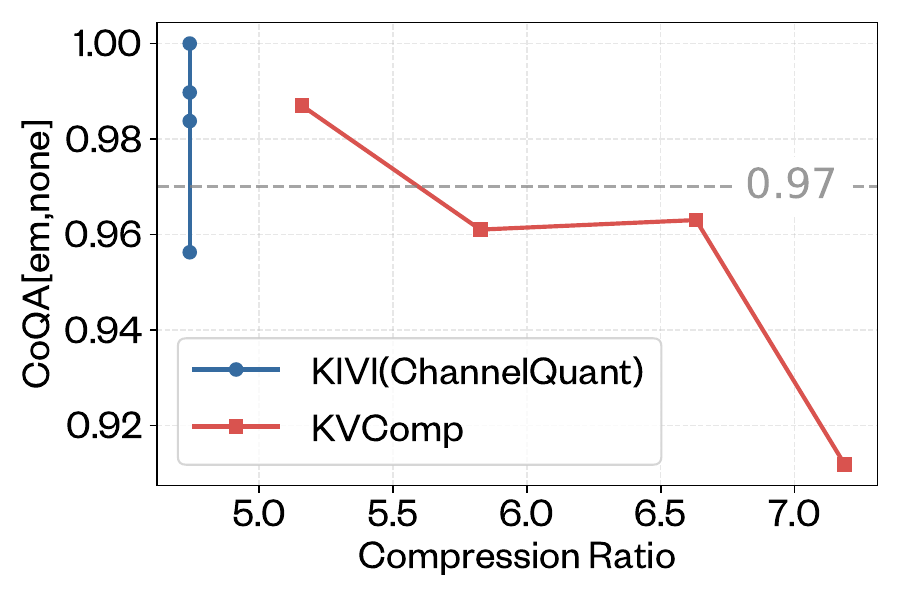}
        \caption{CoQA[em,none]}
        % \label{fig:k_block_q}
    \end{subfigure}
    \hfill
    \begin{subfigure}[b]{0.24\textwidth}
        \centering
        \includegraphics[width=\textwidth]{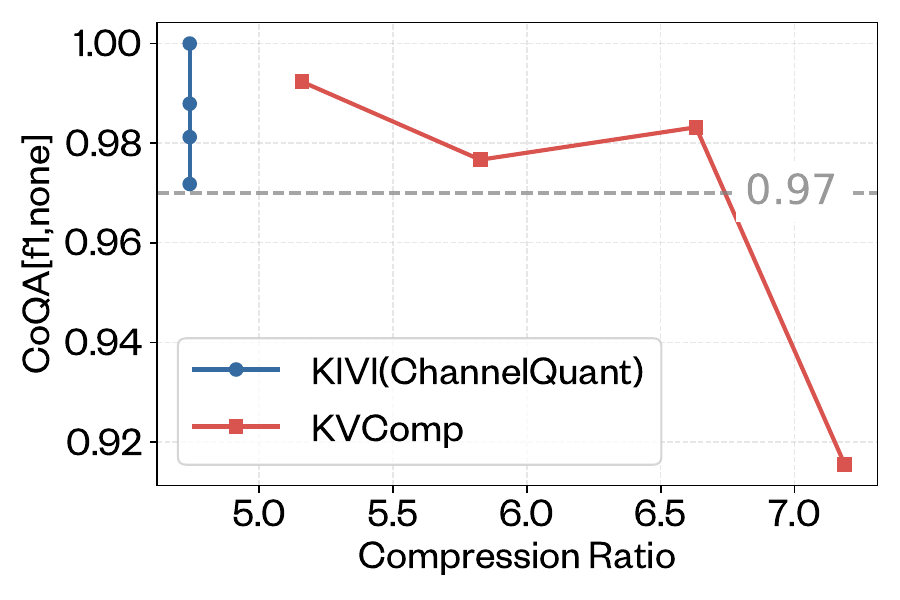}
        \caption{CoQA[f1,none]}
        % \label{fig:hist_k_39}
    \end{subfigure}
    \hfill
    \begin{subfigure}[b]{0.24\textwidth}
        \centering
        \includegraphics[width=\textwidth]{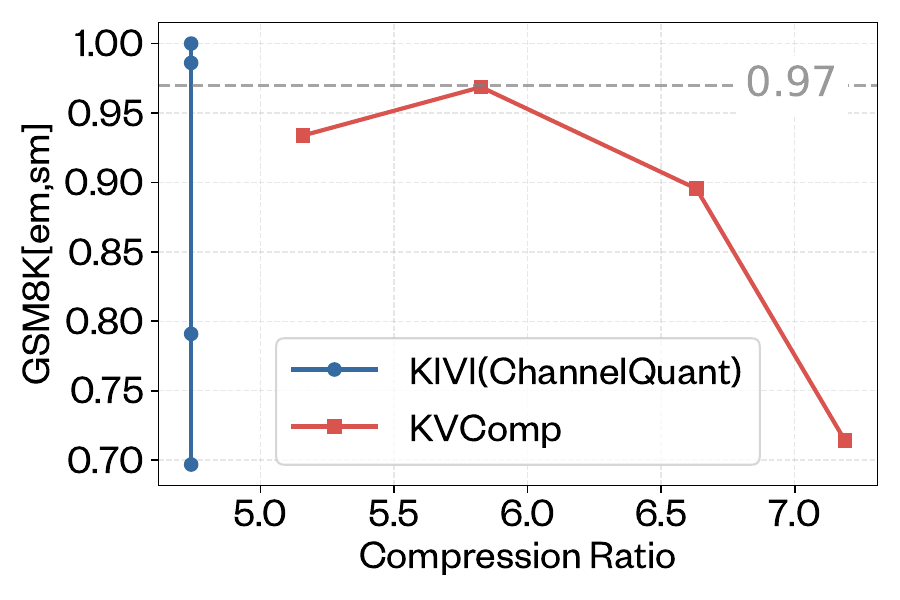}
        \caption{GSM8K[em,sm]}
        % \label{fig:k_block_q}
    \end{subfigure}
    \hfill
    \begin{subfigure}[b]{0.24\textwidth}
        \centering
        \includegraphics[width=\textwidth]{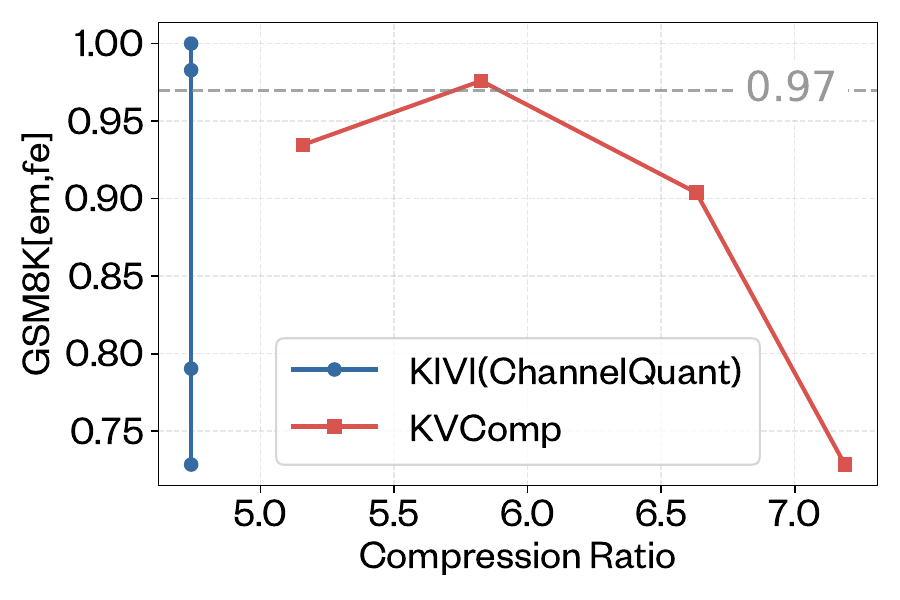}
        \caption{GSM8K[em,fe]}
        % \label{fig:hist_k_39}
    \end{subfigure}
    \vspace{-0.2cm}
    \caption{Accuracy (Nomalized, CoQA [EM, none], GSM8K [EM, SM]) versus relative quantization scale ([K, V]) for Llama2-13B. The reason the KIVI curve is vertical with respect to the x-axis is that, although the quantization scale is varied, the bit-width used to represent each data point can only be an integer; thus, the compression ratio remains unchanged.}
    \Description{Four scatter plots showing accuracy vs compression ratio trade-offs for CoQA-em, CoQA-f1, GSM8K-em-sm, and GSM8K-em-fe benchmarks. Each plot compares KVComp against KIVI, with KIVI showing vertical lines due to discrete bit-width constraints while KVComp shows more flexible compression ratios.}
    \label{fig:k_cr_accu}
\end{figure*}

\subsubsection{Cache-Resident Decompression}

% As previously noted, the most time-intensive operations in our decompression kernel are Huffman decoding and the writeback of decompressed data. 
% While our shared procedure implementation has mitigated branch divergence, it remains essential to further optimize the decompression kernel by eliminating the writeback of decompressed data in the decode kernel. 
We propose to performing decompression directly in cache and consuming the data \emph{in situ}, thereby avoiding the need to write the decompressed output back to global memory. 
Since the KV cache is used for matrix-vector multiplication during the computation of attention weights and outputs, we are able to execute this multiplication operation directly on data stored in cache.
% Moreover, the matrix-vector multiplication kernel is memory-bound, we optimize this step by delivering data to the multiplication operation through shared memory or registers. 
Our 2D blockwise compression and decompression approach, combined with carefully chosen encoding strategies, ensures that decompressed data is immediately available as input for subsequent matrix-vector multiplication operations.

For the K cache with 2D block \texttt{[block\_size, head\_dim]}, each head's Q vector performs a dot product with the corresponding head's K vector. By assigning each thread to process a particular \texttt{[head\_dim]} row within the 2D block, we can directly decompress the vector for a given head in K and immediately accumulate the dot product result using the decompressed data.

For the V cache with 2D block \texttt{[head\_dim, block\_size]}, the attention weights of shape \texttt{[context\_length, head\_num]} are multiplied by the vectors for each head in V. 
We assign each thread to a particular \texttt{[head\_dim]} row of the 2D block, enabling direct decompression of the vectors for specific heads in V and immediate use for dot product accumulation. After fusing the decompression and matrix-vector multiplication kernels, we obtain a matrix of shape \texttt{[context\_length / head\_num, head\_num]}. The final step is a sum reduction, which incurs minimal computational overhead and produces the final result.
\section{Evaluation}
\label{sec:evaluation}

\subsection{Experiment Setup}
\label{subsec:exp_setup}

The experiments are based on two systems. The first cluster features nodes equipped with two Intel Xeon E5-2620v4 processors, offering 16 physical cores in total, along with four NVIDIA Tesla V100 GPUs per node. The second workstation equipped with a single AMD Ryzen Threadripper 7970X CPU (32 physical cores), 256 GB RAM, and two NVIDIA GeForce RTX 4090 GPUs (24 GB each).

We select the model accuracy, compression ratio, and decompression throughput as our metrics.
Specifically, we selected three models (i.e., Llama2 7B, Llama2 13B, and Ministral 8B) and two benchmarks (i.e., CoQA and GSM8K), to evaluate model accuracy in different configurations.
Although our solution can maintains the same accuracy as the standalone quantization methods, we discovered an more suitable quantization granularity for our method that achieves higher compression ratios while preserving model accuracy.
% We selected two benchmarks from the lm-eval framework: CoQA and GSM8K, to measure model accuracy in different compression configurations.
Due to the asymmetric nature of compression and decompression demands in our use case, we focus primarily on decompression throughput. We compare the throughput of our cache-resident decompression method with the original standalone decompressor.

% \subsubsection{Hardware \& Software Environment} 

% All experiments are conducted on a workstation equipped with a single AMD Ryzen Threadripper 7970X CPU (32 physical cores), 256 GB RAM, and two NVIDIA GeForce RTX 4090 GPUs (24 GB each). The system runs Ubuntu 20.04.6 LTS with Linux kernel 5.15.0-136-generic, NVIDIA driver 535.183.01, and CUDA 12.2.

% \subsubsection{Metric Selection} Since our compressor targets LLM inference, a more appropriate metric to measure the error introduced by our compression technique is the impact on model accuracy rather than the conventional error-bound metrics used in compression evaluation. Therefore, our first metric is model accuracy. We also evaluate two standard compression metrics: compression ratio and throughput.

% \textbf{Model Accuracy:} While our compressor maintains the same accuracy as the standalone quantization methods, we discovered an more suitable quantization granularity for our method that achieves higher compression ratios while preserving model accuracy. To evaluate this, we selected two benchmarks from the lm-eval framework: CoQA and GSM8K, to measure model accuracy in different compression configurations.

% \textbf{Compression Ratio:} This metric represents the reduction in data size achieved by our technique.

% \textbf{Decompression Throughput:} Due to the asymmetric nature of compression and decompression demands in our use case, we focus primarily on decompression throughput. We compare the throughput of our cache-resident decompression method with the original standalone decompressor.

We use KIVI~\cite{liu2024kivi} quantization as the baseline for evaluating model accuracy and compression ratio. 
The original data type of the KV cache is \texttt{float16}, existing lossy compressors do not natively support this format. 
Furthermore, no current lossy compressors support fine-grained quantization or handle dynamically growing data structures efficiently.
We also acknowledge that our solution is orthogonal to KV cache pruning and GPU-CPU migration, and can be combined with these methods for further optimization.
For decompression throughput, we use our in-house standalone compressor implementation as the baseline.
We also include raw cuBLAS matrix vector multiplication throughput in our comparison.

% \subsubsection{Baseline Selection} For model accuracy and compression ratio, we use KIVI quantization as our baseline. Since the original datatype of KV cache is float16, there are no existing lossy compressors that natively support this data type. Furthermore, no current lossy compressors support fine-grained quantization or handle dynamically growing data structures efficiently. For decompression throughput, we use our in-house standalone compressor implementation as the baseline. And also after knowing that our single kernel implementation can even accelerate the matrix vector multiplication process, we include PyTorch(use cuBLAS as backend for CUDA device) matrix vector multiplication throughput comparison.

% \subsubsection{Model Selection} To cover the various model classes, we chose Llama2 (7B,13B), Ministral (8B)
% \cb{bo: I removed the phi4 because it is so different from other model}.
% Phi4 (14B)

\subsection{Model Accuracy vs Compression Ratio}

% \subsubsection{K and V standalone test}Based on the transformers Python package, we implemented KVCompCache, which implements necessary interface functions of the Cache class that can be efficiently adopted across all these models. Our compression configuration includes three major parameters: block size (the size of the KV cache truncated from the buffer when it overflows—\cred{xh: is this a dash or the minus sign. I guess it's minus, please confirm.}recent size), the maximum buffer size, and quant relative scale (the quantization relative scale in each quantization block unit). To compare our quantization method with KIVI quantization, we plotted multiple figures where the x-axis represents the compression ratio and the y-axis represents benchmark accuracy.

Based on the HuggingFace \textbf{transformers} Python package, we implemented a \textbf{KVCompCache} class, which provides the required interface functions for the \textbf{Cache} class and can be efficiently integrated with all supported models. 

Our compression configuration includes three key parameters: 
\begin{itemize}
    \item \textbf{Block size}: the size of each block of the K cache that is truncated from the buffer when it overflows 
    % (\cred{xh: is this a dash or a minus sign? I assume it's a minus sign, please confirm.}\cb{bo: removed the dash}recent size).
    \item \textbf{Maximum buffer size}: the upper limit for the cache buffer.
    \item \textbf{Relative quantization scale}: the relative scaling factor applied during quantization within each block.
    Specifically, we define a global relative quantization scale between $[0, 1]$. The actual quantization scale for each block/channel is $rel\_quant\_scale * (max\_value - min\_value)$.
\end{itemize}

% To compare our quantization method with KIVI quantization, we divide the comparison into two parts: \textbf{K comparison} and \textbf{V comparison}. 
% Since our solution uses similar V quantization stratify compared to KIVI, we only assess the V compression ratio between our solution and KIVI.
% Since our method does not modify the V quantization relative to KIVI quantization, there is no need to evaluate model accuracy when adjusting V quantization settings. The only aspect we assess is the V compression ratio of both KIVI quantization and our method, as the only additional step we introduce is Huffman encoding—a lossless encoding that introduces no error to the quantized integers.

\subsubsection{K and V Standalone Test}

First, we conduct standalone accuracy benchmarks for K and V, varying relative quantization scales to identify the turning point at which accuracy drops significantly, shown in Figure~\ref{fig:k_v_separate}.
During this experiment, we only compress K with original V, or compress V with original K to determine their standalone impact to model accuracy.
For K, we conduct both BlockQuant and ChannelQuant for comparison, where ChannelQuant is a similar quantization solution compared to KIVI.
For V, we only conduct TokenQuant as it already align with our decompression design.
We normalize the original model accuracy to 1 and unify all accuracy values accordingly for simplicity. 
We select the largest quantization relative scale at which the model’s accuracy drops by less than 3\%. We choose 3\% because model accuracy drop dramatically if the accuracy drop below 97\% shown in the K and V standalone accuracy test.
Note that although the corresponding $rel\_quant\_scale$ of the turning point of K ChannelQuant is larger than that of K BlockQuant, they are not directly comparible due to their different value range.
We will directly compare the two strategies by their compression ratios in Figure~\ref{fig:k_cr_accu}.
Based on Figure~\ref{fig:k_v_separate}, the accuracy turning point for the K BlockQuant quantization scale is approximately 0.05–0.06. For K ChannelQuant, the turning point can be set around 0.25–0.3. For V TokenQuant, the turning point occurs at approximately 0.15–0.2.

\begin{figure}[tbp]
    \centering
    
    \begin{subfigure}[b]{0.45\textwidth}
        \centering
        \includegraphics[width=\textwidth]{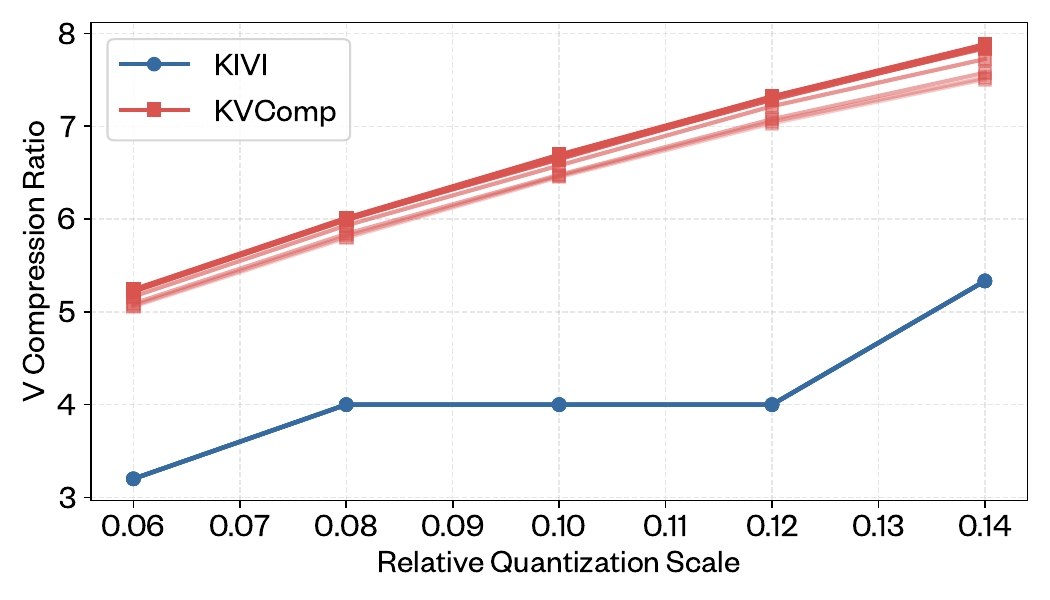}
        \caption{V TokenQuant}
        % \label{fig:k_block_q}
    \end{subfigure}
    \vspace{-0.2cm}
    \caption{V Compression Ratio comparison between KIVI and KVComp across various relative quantization scales. Lines correspond to different context lengths (ctx\_len) [2048-16384], where darker shades indicate larger ctx\_len values.}
    \Description{Line plot comparing V cache compression ratios between KIVI and KVComp methods across different quantization scales. Multiple lines represent various context lengths from 2048 to 16384 tokens, with KVComp showing superior compression ratios compared to KIVI across all configurations.}
    \label{fig:v_eb_cr}
\end{figure}

\subsubsection{K and V Combined Test}

% Figure~\ref{fig:k_v_separate} presents the standalone configurations for K and V. Since keys and values may interact with each other, we conducted additional experiments to evaluate various K–V combinations. 
In this subsection, we introduced a more comprehensive benchmark, GSM8K, to assess the model accuracy drop with both compressed K and V.
We fix the ratio of $rel\_quant\_scale$ of K and $rel\_quant\_scale$ of V based on the turning points from Figure~\ref{fig:k_v_separate}, and conduct our experiments with both K and V compressed, shown in Figure~\ref{fig:k_v_combine}.
% We also fixed the ratio between the relative quantization scales of K and V. Figure~\ref{fig:k_v_combine} shows the results of the K–V combination accuracy tests.
% We note that all K ChannelQuant (i.e., KIVI quantization) failed to achieve a relative quantization scale greater than 0.33 while keeping the accuracy drop below 3\%.
% As shown in the figure, all K ChannelQuant (KIVI) configurations failed to achieve a relative quantization scale greater than 0.33 while keeping the accuracy drop below 3\%. 
% Considering the block size of 64, KIVI requires storing at least an 8-bit quantization minimum integer and a 16-bit floating-point quantization scale for every 64 data points. 
% Thus, if KIVI achieves 2-bit quantization, the average number of bits per data point is \((64 \times 2 + 8 + 16) / 64 = 2.375\), which corresponds to a compression ratio of approximately 6.74. However, since KIVI quantization cannot achieve 2-bit K quantization and requires at least 3 bits per data point, the compression ratio drops to about 4.7—comparable to coarse-grained 4-bit quantization.
We can observe that even with different LLM model size and LLM model structure (i.e., Llama2 7B, Llama2 13B, and Ministral 8B), our solution select consistent $rel\_quant\_scale$ given a benchmark.

\begin{figure}[tbp]
    \centering
    \begin{subfigure}{0.95\linewidth}
        \centering
        \includegraphics[width=1\linewidth]{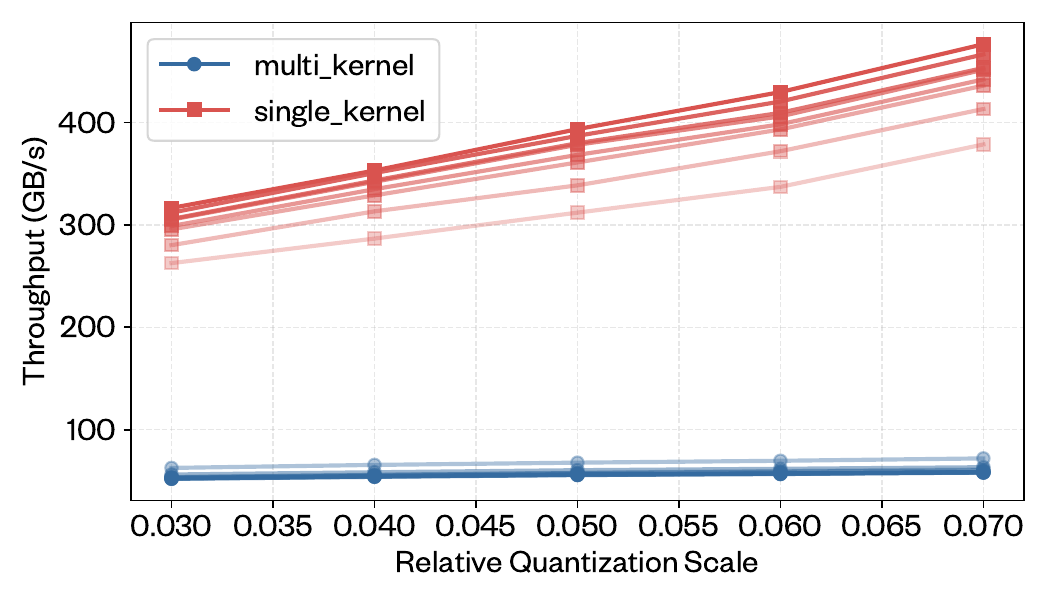}
        \vspace{-0.3cm}
        \caption{K decompression throughput}
        \label{fig:k_multi_single_kernel_time}
    \end{subfigure}
    \vspace{1em}

    \begin{subfigure}{0.95\linewidth}
        \centering
        \includegraphics[width=1\linewidth]{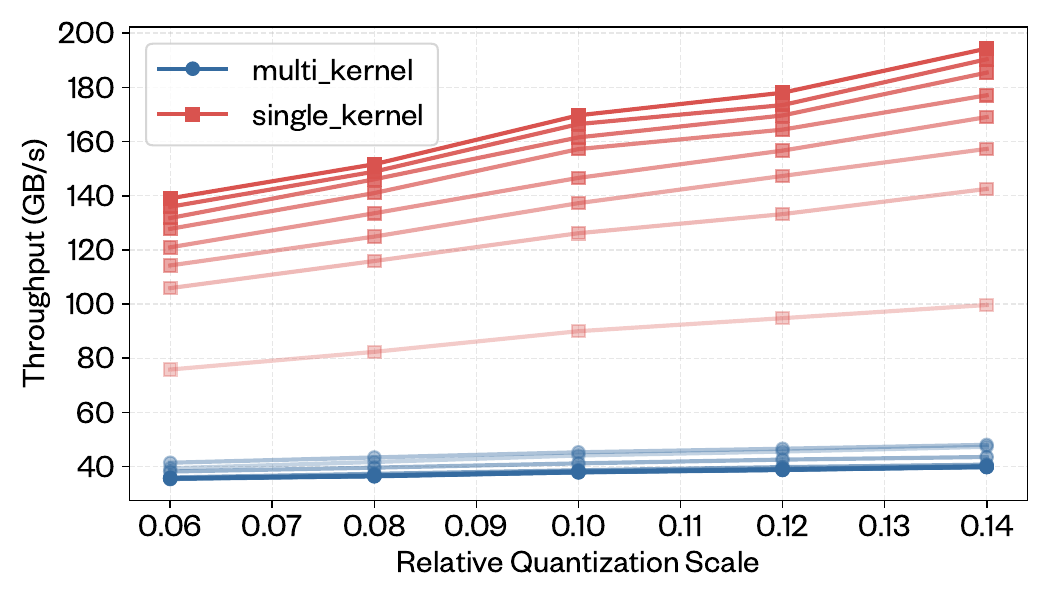}
        \vspace{-0.3cm}
        \caption{V decompression throughput}
        \label{fig:v_multi_single_kernel_time}
    \end{subfigure}
\vspace{-0.2cm}
    \caption{Decompression throughput single kernel (decompress + matrix vector multiplication) vs multi kernels (huffman decode + dequantization + matrix vector multiplication). Lines correspond to different context lengths (\texttt{ctx\_len}) [2048-16384], where darker shades indicate larger \texttt{ctx\_len} values.}
    \Description{Two line plots showing decompression throughput comparisons: top plot for K cache and bottom plot for V cache. Each shows single kernel vs multi-kernel approaches across different batch sizes, with multiple lines representing various context lengths from 2048 to 16384 tokens.}
    \label{fig:multi_single_kernel}
\end{figure}

\begin{figure}[tbp]
    \centering
    \begin{subfigure}{0.95\linewidth}
        \centering
        \includegraphics[width=1\linewidth]{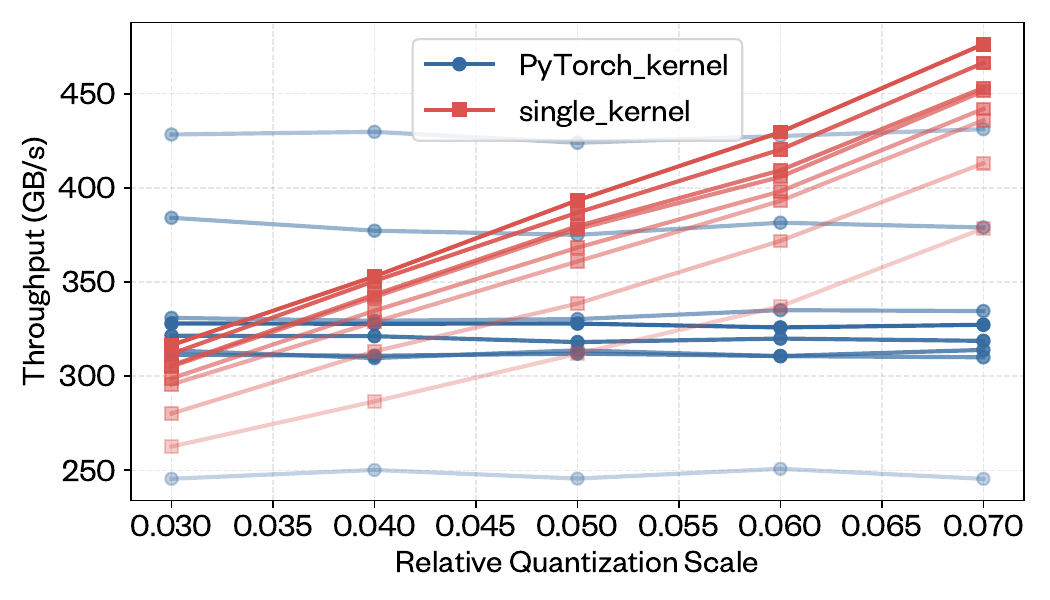}
        \vspace{-0.3cm}
        \caption{K Throughput, KVComp single kernel vs PyTorch kernel}
        \label{fig:k_cublas_single_kernel_time}
    \end{subfigure}
    \vspace{1em}

    \begin{subfigure}{0.95\linewidth}
        \centering
        \includegraphics[width=1\linewidth]{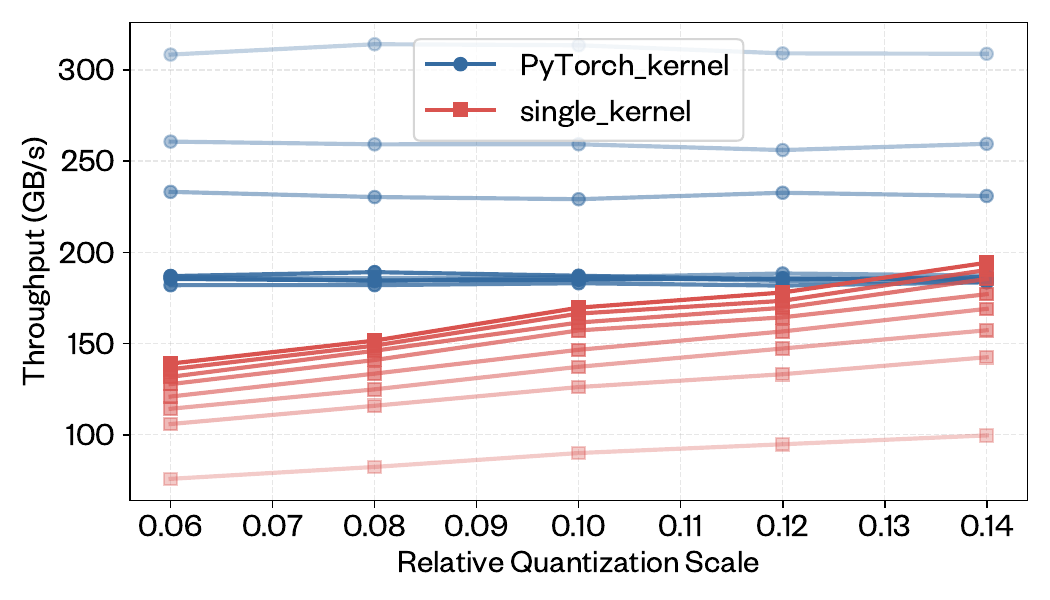}
        \vspace{-0.3cm}
        \caption{V Throughput, KVComp single kernel vs PyTorch kernel}
        \label{fig:v_cublas_single_kernel_time}
    \end{subfigure}
\vspace{-0.2cm}
    \caption{Decompression throughput KVComp single kernel (decompress + matrix vector multiplication) vs PyTorch kernels (matrix vector multiplication only). Lines correspond to different context lengths (\texttt{ctx\_len}) [2048-16384], where darker shades indicate larger \texttt{ctx\_len} values.}
    \Description{Two line plots comparing KVComp single kernel performance against PyTorch kernels: top plot for K cache and bottom plot for V cache. Shows throughput vs batch size with multiple lines for different context lengths, demonstrating KVComp's performance advantages especially at larger context lengths.}
    \label{fig:cublas_single_kernel}
\end{figure}

Furthermore, we conducted experiments to measure the compression ratios with our solution compared to KIVI.
% Once the appropriate relative quantization scale is determined, we conducted experiments to measure the K and V compression ratios. 
For the compression ratio tests, we used text from the WikiText-103-v1 dataset, assembling contexts from its lines and using this text as model input to generate the KV cache.

For K quantization, we established the relationship between compression ratio and accuracy, as shown in Figure~\ref{fig:k_cr_accu}. 
\textcolor{black}{Overall, our method improves the compression ratio by up to 41\% and on average 32\% compared to KIVI with no/minimum accuracy degradation.}
For V quantization, since Token-wise quantization is also used in KIVI, we focused on the compression ratio improvement from our entropy encoding solution, as shown in Figure~\ref{fig:v_eb_cr}. 
We evaluate the compression ratio across different context lengths, ranging from 4,096 to 16,384.
Our method achieves up to 83\% (i.e., Relative quantization scale = 0.12, Context length = 16384) and on average 62\% improvement over KIVI in compression ratio while maintaining the same model accuracy.
Additionally, we observe that our solution achieves a consistent compression ratio across different context length scales, indicating that it scales effectively in large context length scenarios.

\subsection{Inference Computation Throughput}

As discussed in Section~\ref{sec:decompression}, during the decoding stage of LLM inference, each iteration requires compressing the KV cache corresponding to a single input token but decompressing the entire KV cache for the context. This results in a significant imbalance between compression and decompression demands. 
Consequently, we focus on decompression throughput and the corresponding inference computation throughput in our evaluation. 
% Note that the batch size of all the matrix multiplication below are 1.

We use the optimal compression configurations identified in previous analyses, and collect KV cache data during model inference. 
We then record both CUDA kernel execution times and the sizes of original and compressed data from compressing and decompressing KV cache data during inference. 
For standalone decompression kernels, we measured the execution times for Huffman-based decoding, dequantization and matrix-vector multiplication. 
For the cache-resident decompression kernel, we measured the execution time of our fused (i.e., decompression + matrix-vector multiplication) kernel.
First, we compare the kernel times of the multi-kernel and single-kernel implementations; then, we compare our single fused kernel (decompression + matrix vector multiplication) with the cuBLAS-based matrix-vector multiplication kernel.

Figure~\ref{fig:multi_single_kernel} presents the kernel execution times comparison between our single-kernel implementation and the multi-kernel pipeline. As the relative quantization scale increases, the kernel execution time decreases. This is because a larger quantization scale correlate to a higher compression ratio, thus a smaller bits/value KV cache size. Since our decode algorithm processes only one bit per decoding iteration, having fewer bits to decode results in faster decompression. 
% This relationship explains the observed trend in kernel execution times.
Our solution outperforms the multi-kernel implementation in all scenarios.
Additionally, we observe that our solution can provide extremely high throughput of over 400 GB/s for K and over 180 GB/s for V.
Note that this high-throughput single kernel not only include the decompression for KV cache, but also their subsequent matrix multiplication operations.

Next, we compare the kernel execution time of our single-kernel implementation with that of the cuBLAS-based matrix-vector multiplication kernel, as shown in Figure~\ref{fig:cublas_single_kernel}. 
It is important to note that this comparison is not strictly fair, as the cuBLAS kernel performs only matrix multiplication, whereas our kernel includes both decompression and matrix-vector multiplication. However, as the context length increases, our kernel outperforms cuBLAS—even while performing additional work. 
This is primarily because our approach not only keeps data in situ during decompression and computation, avoiding intermediate memory transfers, but also operates on compressed data that loads significantly less data from global memory compared to loading original data by cuBLAS. 
This reduction in data movement becomes substantial at high compression ratios, enabling our fused kernel to surpass cuBLAS in overall performance.
In addition, shown in Figure~\ref{fig:cublas_single_kernel}, our solution scales well with increasing context length. 
We observe a clear performance improvement as the context length grows, which is mainly due to higher hardware utilization when processing larger volumes of data.

\begin{figure}[tbp]
    \centering
    
    \begin{subfigure}[b]{0.45\textwidth}
        \centering
        \includegraphics[width=\textwidth]{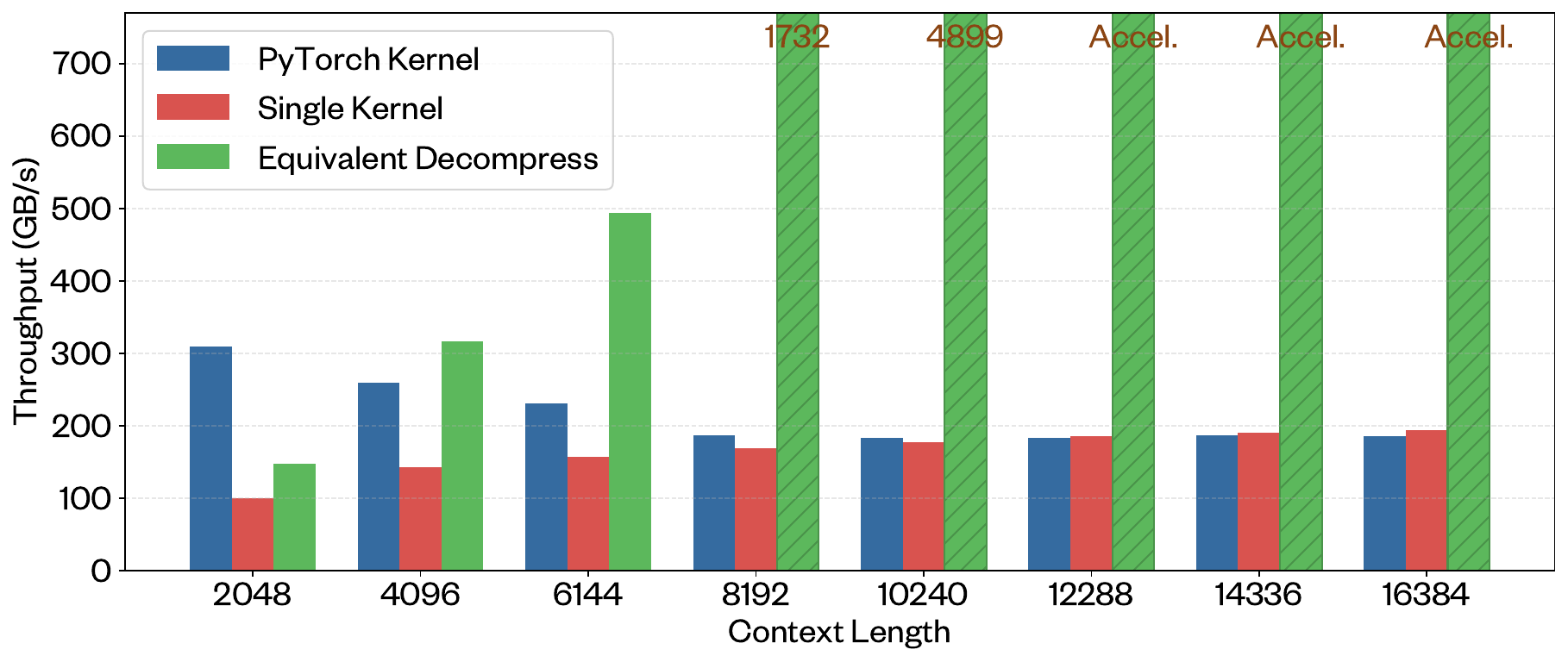}
        % \caption{Context length 16384}
        % \label{fig:k_block_q}
    \end{subfigure}
    \vspace{-0.2cm}
    \caption{V cache, Single kernel (decompress + matrix-vector multiplication), PyTorch kernel, and equivalent decompress. As the context length increases, our single kernel throughput continues to improve, eventually accelerating the matrix-vector multiplication operation, which is typically memory-bound. Thus, lossy compression techniques can not only reduce data size but also enhance computational performance.}
    \Description{Bar chart comparing throughput across different approaches for V cache operations. Shows three categories: single kernel (decompress + matrix-vector multiplication), PyTorch kernel, and equivalent decompress, demonstrating performance improvements with increasing context lengths and the computational benefits of compression.}
    \label{fig:single_cublas_equivalent}
    \vspace{-0.2cm}
\end{figure}

Lastly, we compute the equivalent decompression throughput of our single-kernel design by analyzing its performance difference relative to cuBLAS, as shown in Figure~\ref{fig:single_cublas_equivalent}. We observe that once the context length exceeds 8,192, the equivalent decompression throughput of our kernel becomes significantly higher than that of standalone matrix multiplication. This indicates that in real-world scenarios with long context lengths, our approach introduces negligible performance overhead—or even yields performance gains—while simultaneously providing high compression ratios for substantial memory reduction.

% In conclusion cache resident lossy compression techniques can not only reduce data size but also enhance computational performance.

% \begin{figure}[tbp]
%     \centering
%     \includegraphics[width=0.95\linewidth]{fig/single_cublas_eb/k.pdf}
%     % \vspace{-4mm}
%     \caption{K decompress + mat vec mul single kernel time vs K mat vec mul cublas kernel time. Lines correspond to different context lengths (ctx\_len) [2048-16384], where darker shades indicate larger ctx\_len values.}
%     \label{fig:k_single_vs_cublas}
% \end{figure}

% \begin{figure}[tbp]
%     \centering
%     \begin{subfigure}[]{0.45\textwidth}
%         \centering
%         \includegraphics[width=\textwidth]{fig/single_cublas_eb/k.pdf}
%         % \label{fig:k_block_q}
%     \end{subfigure}
%     \caption{K decompress + mat vec mul single kernel time vs K mat vec mul cublas kernel time. Lines correspond to different context lengths (ctx\_len) [2048-16384], where darker shades indicate larger ctx\_len values.}
%     \label{fig:k_single_vs_cublas}
% \end{figure}

% \begin{figure}[tbp]
%     \centering
    
%     \begin{subfigure}[b]{0.45\textwidth}
%         \centering
%         \includegraphics[width=\textwidth]{fig/single_cublas_eb/v.pdf}
%         % \caption{Context length 16384}
%         % \label{fig:k_block_q}
%     \end{subfigure}
%     \caption{V decompress + mat vec mul single kernel time vs V mat vec mul cublas kernel time. Lines correspond to different context lengths (ctx\_len) [2048-16384], where darker shades indicate larger ctx\_len values.}
%     \label{fig:v_single_vs_cublas}
% \end{figure}

\section{Conclusion and Future Work}
\label{sec:conclusion}

% In this paper, we propose a novel KV cache lossy compression framework. Our framework leverages extra information of quantizaed integer to improve the compression ratio without introducing extra error compare to the quantization method we based on. To meet the performance de
% mands of large language model inference, developed new buffman encoding strategy and huffman decode algorithm deeply integrated with our cache-resident decompress technique. Evaluations show that our solution improves the compression ratio by average 20\% for V cache and 50\% for K cache compared to state-of-the-art quantization method while introduced no extra compression or decompress overhead, sometimes even accelerating the computation.

% In the future, we will explore more efficient way to leverage the higher efficient entropy encoding compression method and improve both the compression ratio and the throughput. And we will also extend our solution to more benchmark and models.

In this paper, we introduce KVComp, a high-performance, LLM-aware lossy compression framework designed to address the growing memory bottleneck posed by the KV cache during LLM inference. By combining fine-grained quantization with GPU-optimized Huffman-based encoding and cache-resident decompression, KVComp significantly reduces the memory footprint of the KV cache while preserving model accuracy and achieving high execution efficiency.
KVComp achieves up to 83\% improvement in compression ratio, with no additional accuracy degradation. 
KVComp also introduce little overhead, or in some cases, faster execution than cuBLAS-based attention kernels due to less data movement.

% In the future, we plan to investigate alternative entropy encoding schemes with better parallelism and compression efficiency, and extend KVComp’s integration to a broader range of LLM architectures and inference engines.

% In this paper, we propose a novel KV cache lossy compression framework. Our approach exploits additional information from quantized integers to enhance compression ratios without introducing any extra error compared to the underlying quantization method. To meet the performance demands of large language model inference, we develop a new GPU huffman encoding parallel strategy and a Huffman decoding algorithm that are tightly integrated with our cache-resident decompression technique. Experimental results demonstrate that our solution achieves, on average, a 62\% higher compression ratio for V cache and a 32\% improvement for K cache over state-of-the-art quantization methods, without incurring any additional decompression overhead—in some cases even accelerating computation.

In the future, we plan to investigate more efficient entropy encoding techniques to further improve both compression ratio and throughput. Additionally, we will extend our solution to a wider range of benchmarks and models.
% \section{Acknowledgement}

% \begin{acks}
% This research includes calculations carried out on HPC resources supported in part by the National Science Foundation through major research instrumentation grant number 1625061 and by the US Army Research Laboratory under contract number W911NF-16-2-0189. We also acknowledge the data resources provided on SDRBench~\cite{sdrbench,Zhao_2020}, which is maintained by Argonne National Laboratory.
% \end{acks}

\newpage
\bibliographystyle{ACM-Reference-Format}
\bibliography{refs}

\end{document}